\documentclass{article}

\usepackage{arxiv}

\usepackage[utf8]{inputenc} 
\usepackage[T1]{fontenc}    
\usepackage{hyperref}       
\usepackage{url}            
\usepackage{booktabs}       
\usepackage{amsfonts}       
\usepackage{nicefrac}       
\usepackage{microtype}      
\usepackage{lipsum}

\usepackage{cite}
\usepackage[margin=0.5cm]{caption}
\usepackage{graphics}
\usepackage{graphicx}
\usepackage{subcaption}
\usepackage{tikz}
\usepackage{float}
\usepackage{amsmath}
\usepackage{amssymb}
\usepackage{verbatim}
\usepackage{multirow}
\usepackage{pgfplots}
\usepackage{mathtools}
\DeclarePairedDelimiter\ceil{\lceil}{\rceil}
\usepackage{amsmath}
\DeclareMathOperator*{\argmax}{argmax} 
\usepackage{pifont}
\usepackage{array, makecell}
\usepackage{url}

\pgfplotsset{compat=newest}

\title{ECOVNet: An Ensemble of Deep Convolutional Neural Networks Based on EfficientNet to Detect COVID-19 From Chest X-rays}

\author{Nihad Karim Chowdhury\\
  Department of Computer Science and Engineering\\
  University of Chittagong\\
  Bangladesh\\ 
  \texttt{nihad@cu.ac.bd}\\
    \And
 Muhammad Ashad Kabir\\
 School of Computing and Mathematics\\
 Charles Sturt University, NSW\\
 Australia\\ 
 \texttt{akabir@csu.edu.au} \\
 \And
 Md. Muhtadir Rahman\\
 Department of Computer Science and Engineering\\University of Chittagong\\
 Bangladesh\\ 
 \texttt{muhtadirrahman@gmail.com} \\
  \And
  Noortaz Rezoana\\
  Department of Computer Science and Engineering\\
  University of Chittagong\\
  Bangladesh\\ 
  \texttt{rezoananoortaz@gmail.com}\\

}

\begin{document}

\maketitle

\begin{abstract}
The perilous COVID-19 disease puts the world in an exotic state of emergency since it overwhelms the global healthcare system.
Consequently,  people's lives are at greater risk due to high mortality as there is no effective vaccine as a precaution against contracting the occurrence of COVID-19 disease.
Also, researchers all over the world are working to develop corresponding vaccines, and at the same time striving to obtain something effective but pragmatic screening technologies, such as medical imaging.
To combat this disease, one of the preeminent screening techniques may be chest X-ray, as it has many historical credentials related to lung diseases that can provide clinical insights.
This paper proposed an ensemble of deep convolutional neural networks (CNN) based on EfficientNet, named ECOVNet, to detect COVID-19 using a large chest X-ray data set.
At first, the open-access large chest X-ray collection is augmented, and then ImageNet pre-trained weights for EfficientNet is transferred with some customized fine-tuning top layers that are trained, followed by an ensemble of model snapshots to classify chest X-rays corresponding to COVID-19, normal, and pneumonia.
The predictions of the model snapshots, which are created during a single training, are combined through two ensemble strategies, i.e., hard ensemble and soft ensemble to ameliorate classification performance and generalization in the related task of classifying chest X-rays.
In addition, a visualization technique is incorporated in the proposed method to highlight areas that distinguish categories, thereby enhancing the understanding of primal components related to COVID-19 infection.
Empirical evaluations show that the ensemble strategy (especially soft ensemble) can significantly improve prediction performance with $97\%$ accuracy, while the precision and recall rate of detecting COVID-19 are both $100\%$.
We believe that ECOVNet can strengthen the resistance to COVID-19 disease, and more broadly, it will propel towards a fully automated and efficacious COVID-19 detection system.

\end{abstract}


\section{Introduction}
Coronavirus disease 2019 (COVID-19) is a contagious disease that was caused by the Severe Acute Respiratory Syndrome Coronavirus 2 (SARS-CoV-2).
The disease was first detected in Wuhan City, Hubei Province, China in December 2019, and was related to contact with a seafood wholesale market and quickly spread to all parts of the world \cite{whowebsitesituationreports}.
The World Health Organization (WHO) promulgated the outbreak of the COVID-19 pandemic on March 11, 2020.
As of September 20, 2020, this perilous virus has not only overwhelmed the world, but also affected millions of lives. So far, there have been $30,675,675$ confirmed COVID-19 cases and $954,417$ confirmed deaths\cite{whowebsite}.
To limit the spread of this infection, all infected countries strive to cover many strategies such as encourage people to maintain social distancing as well as lead hygienic life, enhance the infection screening system through multi-functional testing, seek mass vaccination to reduce the pandemic ahead of time, etc.
The reverse transcriptase-polymerase chain reaction (RT-PCR) is a modular diagnosis method, however, it has certain limitations, such as the accurate detection of suspect patients causes delay since the testing procedures inevitably preserve the strict necessity of conditions at the clinical laboratory\cite{unknown} and false-negative results may lead to greater impact in the prevention and control of the disease\cite{fang_zhang_xie_lin_ying_pang_ji_2020}.

To make up for the shortcomings of RT-PCR testing, researchers around the world are seeking to promote a fast and reliable diagnostic method to detect COVID-19 infection.
The WHO and Wuhan University Zhongnan Hospital respectively issued quick guides\cite{world_health_organization_2020,rapid_advice_2020}, suggesting that in addition to detecting clinical symptoms, chest imaging can also be used to evaluate the disease to diagnose and treat COVID-19.
In \cite{Multinational_Consensus_Statement_2020}, the authors have contributed a prolific guideline for medical practitioners to use chest radiography and computed tomography (CT) to screen and assess the disease progression of COVID-19 cases.
Although CT scans have higher sensitivity, it also has some drawbacks, such as high cost and the need for high doses of radiation during screening, which exposes pregnant women and children to greater radiation risks\cite{davies_wathen_gleeson_2011}.
On the other hand, diagnosis based on chest X-ray appears to be a propitious solution for COVID-19 detection and treatment. 
In\cite{ng_2020}, Ng et al. remarked that COVID-19 infection pulmonary manifestation is immensely delineated by chest X-ray images.
Moreover, in the case of an artificial intelligence (AI)-based disease recognition system, medical practitioners have already emphasized chest X-rays to explore potential symptoms of COVID-19 infection, such as opaque patterns in the lungs\cite{bbcreports}.

The purpose of this study is to ameliorate the accuracy of COVID-19 detection system from chest X-ray images. 
In this context, we contemplate a CNN-based architecture since it is illustrious for its topnotch recognition performance in image classification or detection.
For medical image analysis, higher detection accuracy along with crucial findings is a top aspiration, and in current years, CNN based architectures are comprehensively featured the critical findings related to medical imaging that's why we constructed the proposed architecture with CNN.
In order to achieve the defined purpose, this paper presents a novel CNN based architecture called ECOVNet, exploiting the cutting-edge EfficientNet\cite{tan2019efficientnet} family of CNN models together with ensemble strategies.
The pipeline of the proposed architecture commences with the data augmentation approach, then optimizes and fine-tunes the pre-trained EfficientNet models, creating respective model’s snapshots.
After that, generated model snapshots are integrated into an ensemble, i.e., soft voting and hard voting, to make predictions.
The motivation for using EfficientNets is that they are known for their high accuracy, while being smaller and faster than the best existing CNN architectures.
Moreover, an ensemble technique has proven to be effective in predicting since it produces a lower error rate compared with the prediction of a single model.
Owing to the limited number of COVID-19 images currently available, diagnosing COVID-19 infection is more challenging, thereby investing with a visual explainable approach is applied for further analysis.
In this regard, we use a Gradient-based Class Activation Mapping algorithm, i.e., Grad-CAM\cite{8237336}, providing explanations of the predictions and identifying relevant features associated with COVID-19 infection.
The key contributions of this paper are as follows:
\begin{itemize}
  \item We propose a novel CNN based architecture that includes front-end pre-trained EfficientNets for feature extraction and model snapshots to detect COVID-19 from chest X-rays.
  \item Taking into account the following assumption, the decisions of multiple radiologists are considered in the final prediction, so we propose an ensemble in the proposed architecture to make predictions, thus making a credible and fair evaluation of the system.
   \item We visualize a class activation map through Grad-CAM to explain the prediction as well as to identify the critical regions in the chest X-ray.
  \item Finally, we appraise our architecture with state-of-the-art architectures through empirical observations to highlight the effectiveness of the proposed architecture in detecting COVID-19.
\end{itemize}

The remainder of the paper is arranged as follows: Section \ref{Related Works} discusses related work. 
Section \ref{Methodology} explains the details of the data set and proposed network architecture, as well as its adjustments to the detection of COVID-19 infection.
The results of our experimental evaluation is presented in Section \ref{Experiments and Results}.
Finally, Section \ref{Conclusion} concludes paper and highlights the future work.

\section{Related Works}
\label{Related Works}

Due to the need to identify COVID-19 infections faster, the latest application areas of CNN-based AI systems are booming, which can speed up the analysis of various medical images.
As we all know, a chest X-ray screening is a state-of-the-art technology with historical prospects for image diagnosis systems for detecting pneumonia\cite{DBLP:journals/corr/abs-1711-05225}. In addition, both pneumonia and COVID-19 go through certain infection characteristics (such as the occurrence of severe lung infections). Hence, it has inspired researchers around the world to explore the ability of chest X-rays through various feature extraction methods especially CNN based approaches to detect COVID-19, thus playing a role when the current health care system is exhausted by the pandemic.

An in-depth survey of the application of CNN technology in COVID-19 detection and automatic lung segmentation is explained in\cite{Shoeibi2020AutomatedDA}, with a focus on analysis using X-rays and computed tomography (CT) images.
Halgurd et al.\cite{maghdid2020diagnosing} tested a modified CNN model as well as a modified pre-trained AlexNet\cite{10.5555/2999134.2999257} using their own chest X-ray and CT scan data set while providing accuracy up to 98\% via modified pre-trained model and $94.1\%$ accuracy by using the modified CNN.
Narin et al.\cite{narin2020automatic} achieved the highest accuracy of 98\% by using three pre-trained models with ImageNet\cite{imagenet_cvpr09} weights (such as ResNet50\cite{inproceedings}, Inception v3\cite{inproceedingsinception}, and Inception-ResNet v2\cite{SzegedyIV16}), taking into account two types of images, i.e., COVID-19 and normal images.
A completely new CNN framework named COVID-Net and a large chest x-ray benchmark data set, i.e., COVIDx introduced by Wang et al.\cite{2020arXiv200309871W}. 
The proposed COVID-Net obtained the best test accuracy of $93.3\%$, and studied how COVID-Net uses an interpretability method to predict.
In\cite{Apostolopoulos_2020}, the state-of-the-art CNN architectures (such as VGG19\cite {Simonyan15}, MobileNetV2\cite {Sandler_2018_CVPR}, Inception\cite{inproceedingsinception}, Xception\cite{Chol17Xception}, Inception-ResNet v2\cite{SzegedyIV16}) were trained using transfer learning on ImageNet, and different neural network architectures were used on top of each architecture. The results produced by fine-tuned models demonstrated the proof-of-principle for using CNN with transfer learning to extract radiological features.

The authors of \cite{minaee_kafieh_sonka_yazdani_soufi_2020} prepared a dataset of 5,000 chest x-rays from the publicly available datasets, and a subset of their benchmark utilized to develop a model by fine-tuning four popular pre-trained CNNs (such as ResNet18\cite{inproceedings}, ResNet50, SqueezeNet\cite{i2016squeezenet} and DenseNet121\cite{inproceedingsdense}). The proposed model was evaluated using the remaining images and produced promising results in terms of sensitivity and specificity.
Eduardo et al.\cite{luz2020effective} proposed a new deep learning framework that extends the EfficientNet\cite{tan2019efficientnet} series, which is well known for its excellent prediction performance and fewer computational steps. Their experimental evaluation showed noteworthy classification performance, especially in COVID-19 cases.
Next, Farooq et al.\cite{farooq2020covidresnet} proposed a method called COVID-ResNet, which uses a three-step technique, including gradually adjusting image size, automatic learning rate selection, and then fine-tuning the pre-trained ResNet50 architecture to improve model performance.
A CNN model called DarkCovidNet\cite{Ozturkdeep} proposed for the automatic detection of COVID-19 using chest X-ray images where the proposed method carried out two types of classification, one for binary classification (such as COVID and No-Findings) and another for multi-class (such as COVID, No-Findings and pneumonia) classification. Finally, the authors provided an intuitive explanation through the heat map, so it can assist the radiologist to find the affected area on the chest X-ray.
In another study, Ucar et al.\cite{ucar_korkmaz_2020} proposed a fine-tuned lightweight SqueezeNet, in which the fine-tuned hyper-parameters were obtained through Bayesian optimization, and the performance of the proposed network was superior to some of the existing CNN networks for detecting COVID-19 cases.

Another research\cite{Karim2020DeepCOVIDExplainerEC} proposed an explainable CNN-based method adjusting on a neural ensemble technique followed by highlighting class-discriminating regions named DeepCOVIDExplainer for automatic detection of COVID-19 cases from chest x-ray images.
A study accomplished by Afshar et al.\cite{Afshar2020COVIDCAPSAC} to contribute an efficacious COVID-19 detection system using Capsule Networks(CapsNets)\cite{capsule_em} based CNN architecture, and the authors of this research claimed their system efficacy not only in statistical performance evaluation but also for a lesser number of trainable parameters compared to its counterparts.
Asif et al.\cite{khan_shah_bhat_2020} proposed a model named CoroNet that used Xception architecture pre-trained on ImageNet dataset and trained on their benchmark creating from two publicly available data sets, and carried out two different classification performance measurement, i.e., three and four classes since the overall accuracy of three and four class classification are $95\%$ and $89.6\%$ respectively.
In \cite{mahmud_rahman_fattah_2020}, Mohammad et al. proposed a CNN-based model called CovXNet, which uses depthwise dilated convolution. At first, the model trained with some non-COVID pneumonia images, and further transferred the acquired learning with some additional fine-tuning layers that trained again with a smaller number of chest X-rays related to COVID-19 and other pneumonia cases. As features extracted from different resolutions of X-rays, a stacking algorithm is used in the prediction process, and for multi-class classification, the accuracy of CovXNet is $90.3\%$. 
In another research, Haghanifa et al.\cite{Haghanifar2020COVIDCXNetDC} prepared a new benchmark by amassing the largest public dataset of COVID-19 chest X-ray images from diverse sources and developed a fine-tuned model based on DenseNet121 using CheXNet\cite{rajpurkar2017chexnet} weight while providing statistical performance along with the visual marker to efficaciously localize the critical region of COVID-19 cases.
Another CNN-based modular architecture proposed by Nihad et al.\cite{chowdhury2020pdcovidnet} named PDCOVIDNet (dilated convolution-based COVID-19 detection network), which consists of several blocks (such as a parallel stack of multi-layer filter blocks in a cascade with a classification and visualization block), in the workflow of COVID-19 detection from chest X-ray images. The authors claimed the effectiveness of the model compared with some well known CNN architecture and showed precision and recall of $96.58\%$ and $96.59\%$ respectively in a case of COVID-19 detection.

\begin{table}[!ht]
\begin{minipage}{\textwidth}
\tiny
\centering
\caption{Overview of CNN based architectures for detecting COVID-19 from chest X-rays}
\begin{tabular}{|c|c|c|c|c|c|}
\hline
Method         & Data Source & Architecture                             & Pre-trained Weight & Ensemble & Visualization \\ \hline
Halgurd et al.\cite{maghdid2020diagnosing} & \makecell{ Cohen et al.\cite{joseph2020ai}, \\ BSTI\footnote{https://www.bsti.org.uk/training-and-education/covid-19-bsti-imaging-database/} }   & AlexNet                                  & ImageNet           & No       & No            \\ \hline
Narin et al.\cite{narin2020automatic}   & \makecell{ Cohen et al.\cite{joseph2020ai}, \\ P. Mooney\cite{Mooneyreports} }            & \makecell{ResNet50,\\Inception v3,Inception-ResNet v2} & ImageNet           & No       & No            \\ \hline
Wang et al.\cite{2020arXiv200309871W} & COVIDx\cite{2020arXiv200309871W}&  COVID-Net(Custom CNN)                                        &       ImageNet             &   No       &   GSInquire\cite{lin2019explanations}            \\ \hline
Apostolopoulos et al.\cite{Apostolopoulos_2020} &  \makecell{ Cohen et al.\cite{joseph2020ai}, Kermany et al.\cite{kermany_2018},\\RSNA\cite{northamerica_2019}, SIRM\cite{SIRMreports}}           &  \makecell{Xception,Inception-ResNet v2, \\VGG19, MobileNet v2,Inception}&     ImageNet  &     No     &     No          \\ \hline
 Shervin et al.\cite{minaee_kafieh_sonka_yazdani_soufi_2020}& COVID-Xray-5k dataset\cite{minaee_kafieh_sonka_yazdani_soufi_2020}            &  \makecell{ResNet18, ResNet50,\\ SqueezeNet, DenseNet121}                                       &   ImageNet                 &    No      &  \makecell{Heat Map,\\Radiologist-Marked}             \\ \hline
  Eduardo et al.\cite{luz2020effective} & COVIDx  &  EfficientNet &    Imagenet   &   No       & Heat Map            \\ \hline
 Farooq et al.\cite{farooq2020covidresnet}  &  COVIDx &  ResNet50  &    ImageNet &  No        &      No         \\ \hline
  Ozturk et al.\cite{Ozturkdeep}&  \makecell{Cohen et al.\cite{joseph2020ai},\\ NIH Chest X-ray dataset\cite{wang_peng_lu_lu_bagheri_summers_2017}}         &  DarkCovidNet(Custom CNN)                                        &     No               &     No     &  \makecell{Grad-CAM,\\Radiologist-Marked}             \\ \hline
   Ucar et al.\cite{ucar_korkmaz_2020}  & COVIDx  & SqueezeNet  &     ImageNet   &    No      &      Heat Map         \\ \hline
    Karim et al.\cite{Karim2020DeepCOVIDExplainerEC}          &  COVIDx           &  \makecell{VGG16,VGG19,ResNet18,\\ ResNet34, DenseNet161, DenseNet201}                                        &  Imagenet                  &  Yes        &   \makecell{Grad-CAM,\\ Grad-CAM++\cite{8354201},LRP\cite{Bach2015OnPE}}            \\ \hline
   Afshar et al.\cite{Afshar2020COVIDCAPSAC}&   COVIDx          & Capsule Networks(CapsNets)                                         &    NIH Chest X-ray dataset                &   No       &      No         \\ \hline
  Asif et al.\cite{khan_shah_bhat_2020} & Cohen et al.\cite{joseph2020ai}, P. Mooney\cite{Mooneyreports}            &  Xception   &  Imagenet     &  No         & No               \\ \hline
  Mohammad et al.\cite{mahmud_rahman_fattah_2020}   &    \makecell{Mendeley Data, V2\cite{Kermany2018LabeledOC}, \\ 305 COVID-19 images }        &             CovXNet(Custom CNN)                             &   Non-COVID X-rays                 &  Yes        &    Grad-CAM           \\ \hline
Haghanifa et al.\cite{Haghanifar2020COVIDCXNetDC}   & COVID-19 Data Collection\footnote{ https://github.com/armiro/COVID-CXNet}            &     DenseNet121                                     &  ImageNet, CheXNet\cite{DBLP:journals/corr/abs-1711-05225}                  &    No      &   Grad-CAM,LIME\cite{Why_Should_I_Trust_You_2016}            \\ \hline
 Nihad et al.\cite{chowdhury2020pdcovidnet} & COVID-19 Radiography Database\cite{chowdhury2020ai}            &   PDCOVIDNet(Custom CNN)    &     No &  No        & Grad-CAM, Grad-CAM++              \\ \hline
  \end{tabular}
\end{minipage}
\label{Summarized Related Work}
\end{table}

It can be seen from the literature review that most methods make prediction decisions based on the output of a single model rather than on ensemble, but few methods\cite{Karim2020DeepCOVIDExplainerEC,mahmud_rahman_fattah_2020} rely on an ensemble.
As we have seen, the ensemble brings a benefit, that is, it can reduce prediction errors, thus making the model more versatile.
One of the previous studies used ensemble on heterogeneous models, i.e., VGG19, ResNet18, and DenseNet161 in \cite{Karim2020DeepCOVIDExplainerEC}, but that approach has some limitations such as that each model requires a separate training session, and an individual model suffers from training many parameters.
Another method\cite{mahmud_rahman_fattah_2020} is to perform an ensemble on a single model, but uses various image resolutions, and for each image resolution, it creates a separate model and stacks it for prediction, which incurs computational overhead.
Contrary to the ensemble, an advanced custom CNN architecture, COVID-Net\cite{2020arXiv200309871W}, was implemented and tested using a large COVID-19 benchmark, but due to the large number of parameters, the computational overhead of this model is high.
To address the aforementioned problems, we use a lightweight but effective model EfficientNet since it is 8.4 times smaller and 6.1 times faster than the best existing CNN\cite{tan2019efficientnet}. 
Also, to extenuate the limitation related to the computational cost of training multiple deep learning models for ensemble prediction, we force large changes in model weights through the recurrent learning rate, creating model snapshots in the same training, and further apply an ensemble to make the proposed architecture more robust.


\section{Methodology}
\label{Methodology}

In this section, we briefly discuss our approach. First, we will precede the benchmark data set and data augmentation strategy used in the proposed architecture. Next, we will outline the proposed ECOVNet architecture, including network construction using a pre-trained EfficientNet and training methods, and then model ensemble strategies. Finally, to make disease detection more acceptable, we will integrate decision visualizations to highlight pivotal facts with visual markers.

\subsection{Dataset}
\label{Dataset}

\begin{figure}[!ht]
  \centering
  \begin{subfigure}[b]{0.3\linewidth}
    \includegraphics[width=\linewidth]{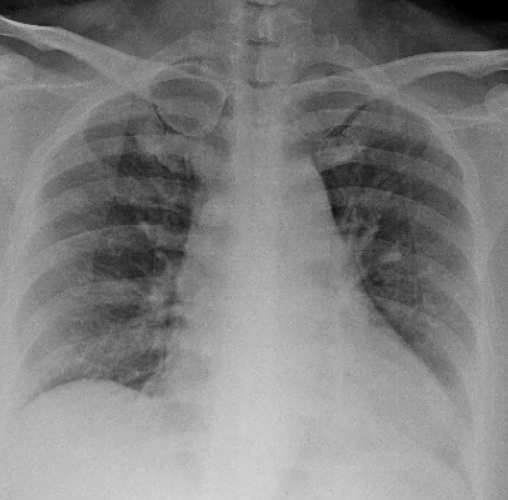}
    \captionsetup{justification=centering}
    \caption{COVID-19}
  \end{subfigure}
  \begin{subfigure}[b]{0.3\linewidth}
    \includegraphics[width=\linewidth]{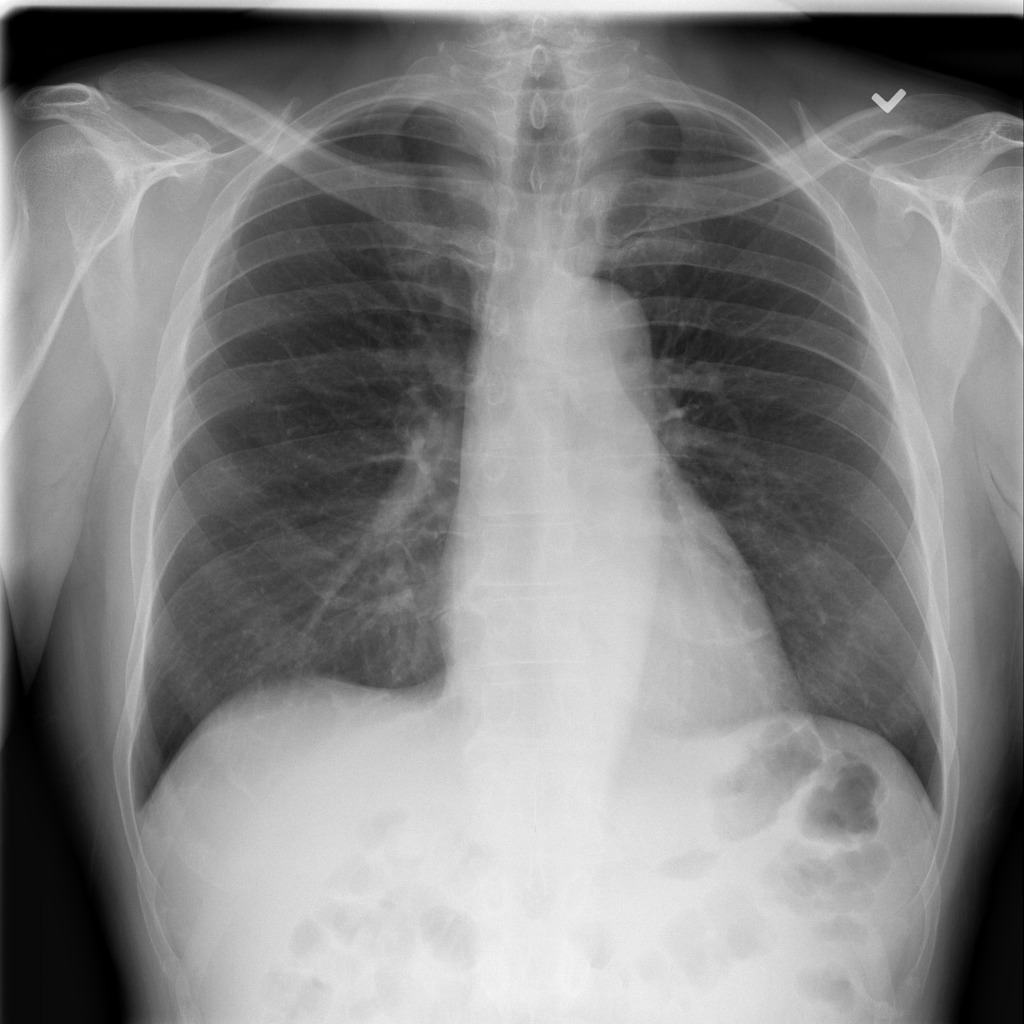}
    \captionsetup{justification=centering}
    \caption{Normal}
  \end{subfigure}
  \begin{subfigure}[b]{0.3\linewidth}
    \includegraphics[width=\linewidth]{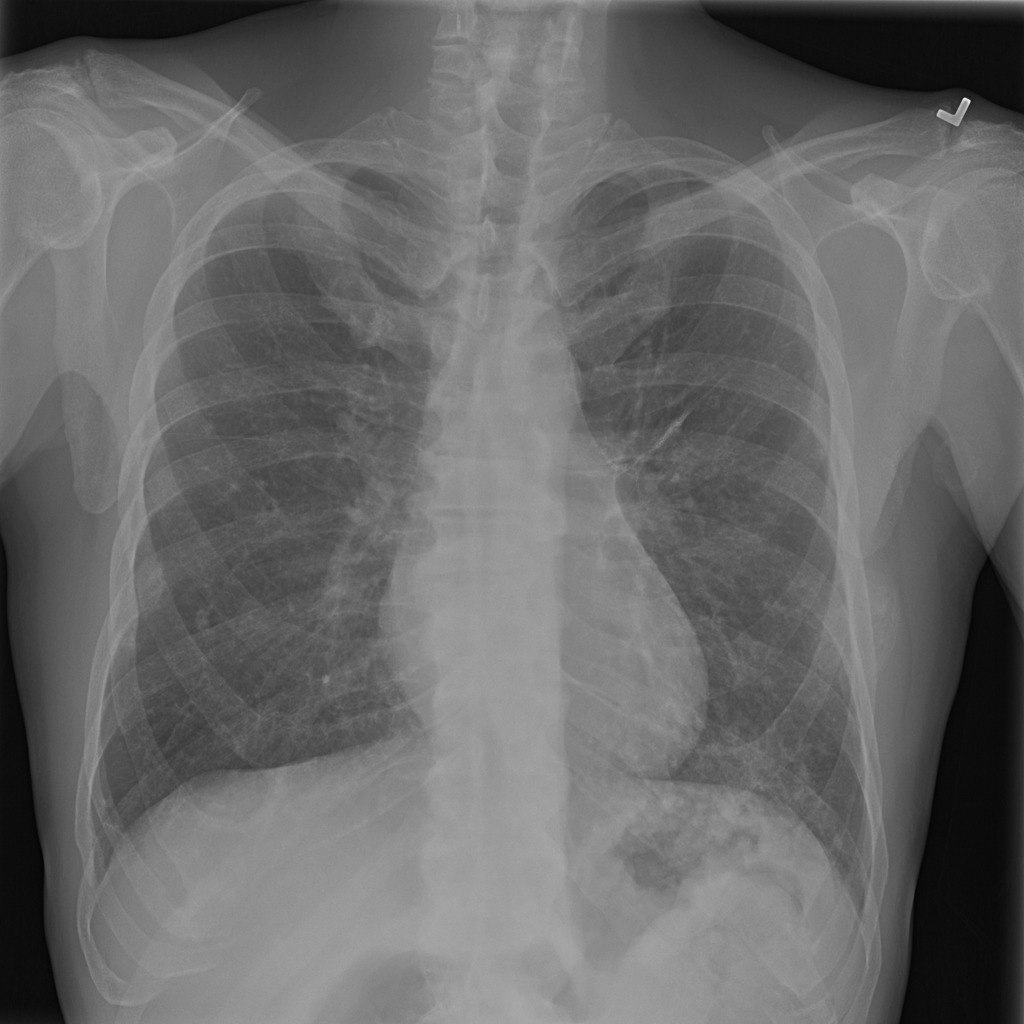}
    \captionsetup{justification=centering}
    \caption{Pneumonia}
  \end{subfigure}
  \captionsetup{justification=centering}
  \caption{Some image labels available in the benchmark dataset \cite{2020arXiv200309871W}}
  \label{fig:benchmark}
\end{figure}

In this sub-section, we concisely inaugurate the benchmark data set, named COVIDx\cite{2020arXiv200309871W}, that used in our experiment. To the best of our knowledge, this data set is one of the largest open-access benchmark data set for the number of COVID-19 infection cases, and the total number of 14,914 images for training and 1,579 images for testing, comprising three categories of COVID-19, normal and pneumonia\footnote{Access on July 17, 2020}.
Figure \ref{fig:benchmark} shows sample images from the benchmark dataset, including COVID-19, normal and pneumonia. Table \ref {image partition} depicts the distribution of images in training and testing sets.
To generate the COVIDx, the authors\cite{2020arXiv200309871W} used five different publicly accessible data repositories:
\begin{itemize}
\item From COVID-19 Image Data Collection\cite{joseph2020ai}, they gathered non-COVID19 pneumonia and COVID-19 cases. 
\item The Figure 1 COVID-19 Chest X-ray Dataset\cite{chung_2020} Initiative utilized for COVID-19 cases.
\item ActualMed COVID-19 Chest X-ray Dataset Initiative\cite{chung_2020_one} selected for COVID-19 cases. 
\item Radiological Society of North America (RSNA) Pneumonia Detection Challenge dataset\cite{northamerica_2019} employed for normal and non-COVID19 pneumonia cases.
\item COVID-19 radiography database\cite{chowdhury2020ai} managed for COVID-19 cases.
\end{itemize}

\begin{table}[!ht]
\centering
\caption{Distribution of images in training and testing sets\cite{2020arXiv200309871W}}
\captionsetup{justification=centering}
      \begin{tabular}{ccccc}
        \hline
        Category  & COVID-19  & Normal   & Pneumonia & Total\\ \hline \hline
        Training   & 489        & 7,966        & 5,459                & 14,914 \\
        Testing & 100        & 885        & 594                 & 1,579 \\ \hline
      \end{tabular}
\label{image partition}
\end{table}

\subsection{Data Augmentation}
\label{Data Augmentation}

Data augmentation is a process performed in time during the training process to expand the training set. As long as the semantic information of an image is preserved, the transformation of the images in the training data set can be used for data augmentation.
Using data augmentation, the performance of the model can be improved by solving the problem of overfitting thus greatly improve inductive reasoning.
Although the CNN model has properties such as partial translation-invariant, augmentation strategies i.e., translated images can often considerably enhance generalization capabilities \cite{goodfellow_bengio_courville_2016}.
Data augmentation strategies provide various alternatives, each of which has the advantage of interpreting images in multiple ways to present important features, thereby improving the performance of the model.
We have considered the following parameters: horizontal flip, rotation, shear, and zoom for augmentation during the training process.


\subsection{Proposed ECOVNet Architecture}
\label{Proposed Network Architecture}

\begin{figure}[!ht]
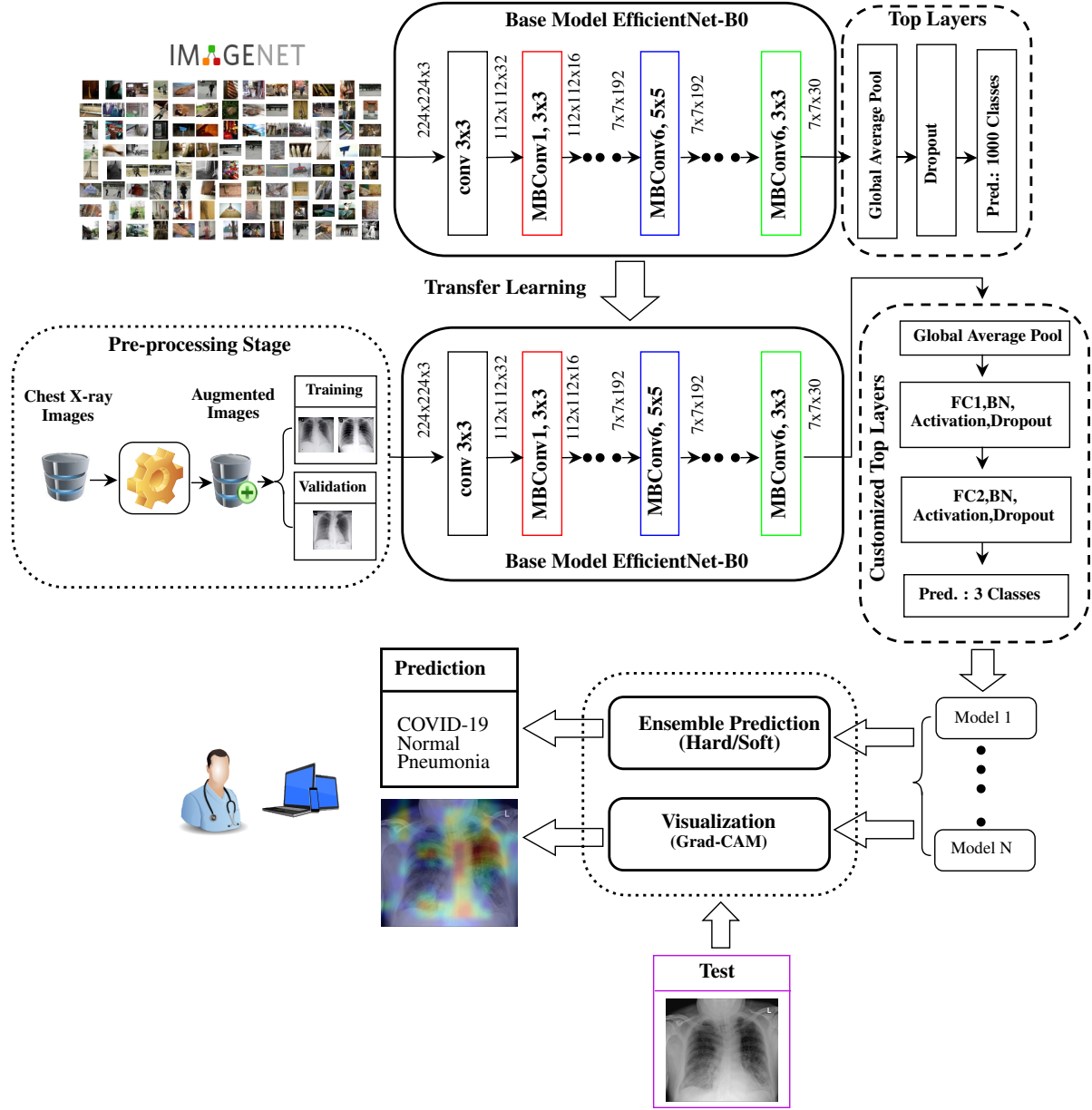

\centering
\resizebox{16cm}{!}{

\tikzset{every picture/.style={line width=0.75pt}} 



}
\caption{Graphical representation of the proposed ECOVNet architecture}
\label{fig:Proposed network architecture}

\end{figure}

In this section, we will briefly describe the proposed ECOVNet architecture.
After augmenting the COVIDx dataset, we used pre-trained EfficientNet as a feature extractor.
This step ensures that the pre-trained EfficientNet can extract and learn useful chest X-ray features, and can generalize it well.
Indeed, EfficientNets are an order of models that are obtained from a base model, i.e., EfficientNet-B0. 
In the proposed architecture, we demonstrated EfficientNet-B0, however, during the experimental evaluation, we considered other models.
The output features from the pre-trained EfficientNet fed to our proposed custom top layers through two fully connected layers, which are respectively integrated with batch normalization, activation, and dropout.
We generated several snapshots in a training session, and then combined their predictions with an ensemble prediction.
At the same time, the visualization approach, which can qualitatively analyze the relationship between input examples and model predictions, was incorporated into the following part of the proposed model.
Fig.\ref {fig:Proposed network architecture} shows a graphical presentation of the proposed ECOVNet architecture using a pre-trained EfficientNet. 

\subsubsection{Pre-trained Efficientnet Feature Extraction}
\label{Pre-trained Efficientnet Feature Extractor}

EfficientNets are a series of models (namely EfficientNet-B0 to B7) that are derived from the baseline network (often called EfficientNet-B0) by scale it up.
The advantages of EfficientNets are reflected in two aspects, namely, it not only provides higher accuracy, but also ameliorates the effectiveness of the model by reducing parameters and FLOPS(Floating Point Operations Per Second).
By adopting a compound scaling method in all dimensions of the network, i.e., width, depth, and resolution, EfficientNets have pulled attention due to its supremacy in prediction performance.
Mention that, width refers to the number of channels in any layer, depth relates to the number of layers in CNN, and resolution associates with the size of the image.
The intuition of using compound scaling is that scaling any dimension of the network (such as width, depth, or image resolution) can increase accuracy, but for larger models, the accuracy gain will decrease.
To scale the dimensions of the network systematically, compound scaling uses a compound coefficient that controls how many more resources are functional for model scaling, and the dimensions are scaled by the compound coefficient in the following way\cite{tan2019efficientnet}:

\begin{equation}
\begin{aligned}
\text{depth:}\,d &= \alpha^\phi \\
\text{width:}\,w &= \beta^\phi \\
\text{resolution:}\,r &= \gamma^\phi \\
&\text{s.t.}\; \alpha.\beta^2.\gamma^2\approx 2 \\
&\alpha \geq 1, \beta \geq 1, \gamma \geq 1 
\label{compoundscaling}
\end{aligned}
\end{equation}

where $\phi$ is the compound coefficient, and $\alpha$, $\beta$, and $\gamma$ are the scaling coefficients of each dimension that can be fixed by a grid search.
After determining the scaling coefficients, these coefficients are applied to the baseline network (EfficientNet-B0) for scaling to obtain the desired target model size.
For instance, in the case of EfficientNet-B0, when $\phi=1$ is set, the optimal values are yielded using a grid search, i.e., $\alpha=1.2$, $\beta=1.1$, and $\gamma=1.15$, under the constraint of $\alpha.\beta^2.\gamma^2\approx 2$\cite{tan2019efficientnet}.
By changing the value of $\phi$ in Equation \ref{compoundscaling}, EfficientNet-B0 can be scaled up to obtain EfficientNet-B1 to B7.

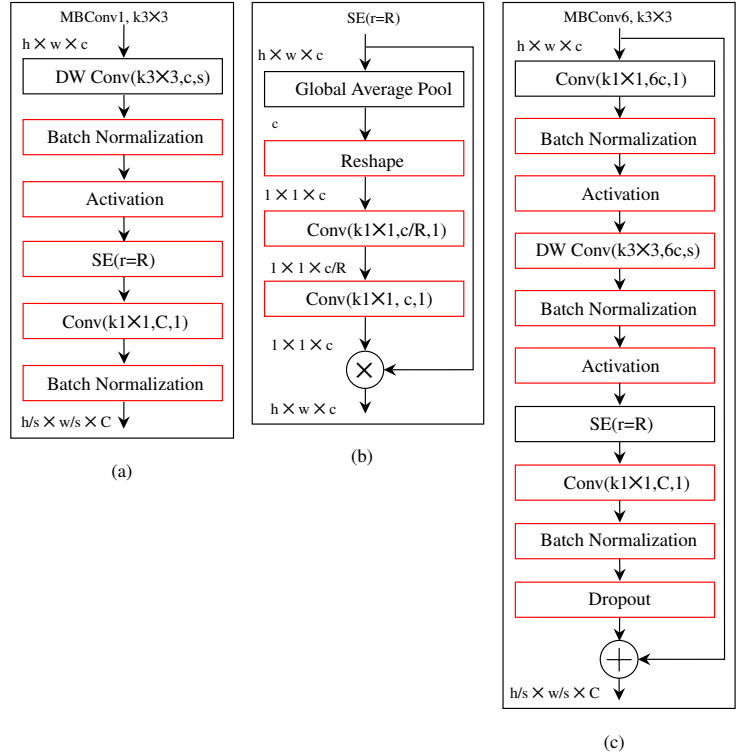
\begin{figure}[!ht]
\centering
\resizebox{9.8cm}{!}{

\tikzset{every picture/.style={line width=0.75pt}} 

\begin{tikzpicture}[x=0.75pt,y=0.75pt,yscale=-1,xscale=1]

\draw   (5.5,0) -- (195.5,0) -- (195.5,372) -- (5.5,372) -- cycle ;
\draw   (16,48) -- (186,48) -- (186,78) -- (16,78) -- cycle ;
\draw  [color={rgb, 255:red, 255; green, 0; blue, 0 }  ,draw opacity=1 ] (16,100) -- (186,100) -- (186,130) -- (16,130) -- cycle ;
\draw  [color={rgb, 255:red, 255; green, 0; blue, 0 }  ,draw opacity=1 ] (16,153) -- (186,153) -- (186,183) -- (16,183) -- cycle ;
\draw  [color={rgb, 255:red, 255; green, 0; blue, 0 }  ,draw opacity=1 ] (16,204) -- (186,204) -- (186,234) -- (16,234) -- cycle ;
\draw  [color={rgb, 255:red, 255; green, 0; blue, 0 }  ,draw opacity=1 ] (16,257) -- (186,257) -- (186,287) -- (16,287) -- cycle ;
\draw  [color={rgb, 255:red, 255; green, 0; blue, 0 }  ,draw opacity=1 ] (16,310) -- (186,310) -- (186,340) -- (16,340) -- cycle ;
\draw   (211.5,0) -- (409,0) -- (409,360) -- (211.5,360) -- cycle ;
\draw   (222,59) -- (392,59) -- (392,89) -- (222,89) -- cycle ;
\draw  [color={rgb, 255:red, 255; green, 0; blue, 0 }  ,draw opacity=1 ] (222,118) -- (392,118) -- (392,148) -- (222,148) -- cycle ;
\draw  [color={rgb, 255:red, 255; green, 0; blue, 0 }  ,draw opacity=1 ] (222,178) -- (392,178) -- (392,208) -- (222,208) -- cycle ;
\draw  [color={rgb, 255:red, 255; green, 0; blue, 0 }  ,draw opacity=1 ] (222,238) -- (392,238) -- (392,268) -- (222,268) -- cycle ;
\draw   (425.5,0) -- (623.5,0) -- (623.5,602.81) -- (425.5,602.81) -- cycle ;
\draw   (436,50) -- (606,50) -- (606,80) -- (436,80) -- cycle ;
\draw  [color={rgb, 255:red, 255; green, 0; blue, 0 }  ,draw opacity=1 ] (436,99) -- (606,99) -- (606,129) -- (436,129) -- cycle ;
\draw  [color={rgb, 255:red, 255; green, 0; blue, 0 }  ,draw opacity=1 ] (436,148) -- (606,148) -- (606,178) -- (436,178) -- cycle ;
\draw  [color={rgb, 255:red, 255; green, 0; blue, 0 }  ,draw opacity=1 ] (436,197) -- (606,197) -- (606,227) -- (436,227) -- cycle ;
\draw  [color={rgb, 255:red, 255; green, 0; blue, 0 }  ,draw opacity=1 ] (436,246) -- (606,246) -- (606,276) -- (436,276) -- cycle ;
\draw  [color={rgb, 255:red, 255; green, 0; blue, 0 }  ,draw opacity=1 ] (436,295) -- (606,295) -- (606,325) -- (436,325) -- cycle ;
\draw  [color={rgb, 255:red, 255; green, 0; blue, 0 }  ,draw opacity=1 ] (436,395) -- (606,395) -- (606,425) -- (436,425) -- cycle ;
\draw  [color={rgb, 255:red, 255; green, 0; blue, 0 }  ,draw opacity=1 ] (436,445) -- (606,445) -- (606,475) -- (436,475) -- cycle ;
\draw  [color={rgb, 255:red, 255; green, 0; blue, 0 }  ,draw opacity=1 ] (436,495) -- (606,495) -- (606,525) -- (436,525) -- cycle ;
\draw   (436,345) -- (606,345) -- (606,375) -- (436,375) -- cycle ;
\draw    (102,18.78) -- (101.86,44.33) ;
\draw [shift={(101.84,47.33)}, rotate = 270.32] [fill={rgb, 255:red, 0; green, 0; blue, 0 }  ][line width=0.08]  [draw opacity=0] (10.72,-5.15) -- (0,0) -- (10.72,5.15) -- (7.12,0) -- cycle    ;
\draw    (102,79) -- (102.17,96.05) ;
\draw [shift={(102.2,99.05)}, rotate = 269.43] [fill={rgb, 255:red, 0; green, 0; blue, 0 }  ][line width=0.08]  [draw opacity=0] (10.72,-5.15) -- (0,0) -- (10.72,5.15) -- (7.12,0) -- cycle    ;
\draw    (102,341) -- (101.86,361.02) ;
\draw [shift={(101.84,364.02)}, rotate = 270.4] [fill={rgb, 255:red, 0; green, 0; blue, 0 }  ][line width=0.08]  [draw opacity=0] (10.72,-5.15) -- (0,0) -- (10.72,5.15) -- (7.12,0) -- cycle    ;
\draw    (102,130.09) -- (102.17,148.65) ;
\draw [shift={(102.2,151.65)}, rotate = 269.47] [fill={rgb, 255:red, 0; green, 0; blue, 0 }  ][line width=0.08]  [draw opacity=0] (10.72,-5.15) -- (0,0) -- (10.72,5.15) -- (7.12,0) -- cycle    ;
\draw    (102,183.09) -- (102.17,201.65) ;
\draw [shift={(102.2,204.65)}, rotate = 269.47] [fill={rgb, 255:red, 0; green, 0; blue, 0 }  ][line width=0.08]  [draw opacity=0] (10.72,-5.15) -- (0,0) -- (10.72,5.15) -- (7.12,0) -- cycle    ;
\draw    (102,234.09) -- (102.17,252.65) ;
\draw [shift={(102.2,255.65)}, rotate = 269.47] [fill={rgb, 255:red, 0; green, 0; blue, 0 }  ][line width=0.08]  [draw opacity=0] (10.72,-5.15) -- (0,0) -- (10.72,5.15) -- (7.12,0) -- cycle    ;
\draw    (102,287.09) -- (102.17,305.65) ;
\draw [shift={(102.2,308.65)}, rotate = 269.47] [fill={rgb, 255:red, 0; green, 0; blue, 0 }  ][line width=0.08]  [draw opacity=0] (10.72,-5.15) -- (0,0) -- (10.72,5.15) -- (7.12,0) -- cycle    ;
\draw    (308.49,38) -- (400,38) -- (400,313.6) -- (326.67,313.65) ;
\draw [shift={(323.67,313.65)}, rotate = 359.96000000000004] [fill={rgb, 255:red, 0; green, 0; blue, 0 }  ][line width=0.08]  [draw opacity=0] (10.72,-5.15) -- (0,0) -- (10.72,5.15) -- (7.12,0) -- cycle    ;
\draw    (308,27) -- (307.86,55.33) ;
\draw [shift={(307.84,58.33)}, rotate = 270.29] [fill={rgb, 255:red, 0; green, 0; blue, 0 }  ][line width=0.08]  [draw opacity=0] (10.72,-5.15) -- (0,0) -- (10.72,5.15) -- (7.12,0) -- cycle    ;
\draw    (308,89.78) -- (307.86,115.33) ;
\draw [shift={(307.84,118.33)}, rotate = 270.32] [fill={rgb, 255:red, 0; green, 0; blue, 0 }  ][line width=0.08]  [draw opacity=0] (10.72,-5.15) -- (0,0) -- (10.72,5.15) -- (7.12,0) -- cycle    ;
\draw    (308,148.78) -- (307.86,174.33) ;
\draw [shift={(307.84,177.33)}, rotate = 270.32] [fill={rgb, 255:red, 0; green, 0; blue, 0 }  ][line width=0.08]  [draw opacity=0] (10.72,-5.15) -- (0,0) -- (10.72,5.15) -- (7.12,0) -- cycle    ;
\draw    (308,208.78) -- (307.86,234.33) ;
\draw [shift={(307.84,237.33)}, rotate = 270.32] [fill={rgb, 255:red, 0; green, 0; blue, 0 }  ][line width=0.08]  [draw opacity=0] (10.72,-5.15) -- (0,0) -- (10.72,5.15) -- (7.12,0) -- cycle    ;
\draw   (291.93,313.65) .. controls (291.93,304.88) and (299.03,297.78) .. (307.8,297.78) .. controls (316.56,297.78) and (323.67,304.88) .. (323.67,313.65) .. controls (323.67,322.41) and (316.56,329.52) .. (307.8,329.52) .. controls (299.03,329.52) and (291.93,322.41) .. (291.93,313.65) -- cycle ;
\draw    (308,268.78) -- (307.86,294.33) ;
\draw [shift={(307.84,297.33)}, rotate = 270.32] [fill={rgb, 255:red, 0; green, 0; blue, 0 }  ][line width=0.08]  [draw opacity=0] (10.72,-5.15) -- (0,0) -- (10.72,5.15) -- (7.12,0) -- cycle    ;
\draw    (308,329.8) -- (308,348.07) ;
\draw [shift={(308,351.07)}, rotate = 270] [fill={rgb, 255:red, 0; green, 0; blue, 0 }  ][line width=0.08]  [draw opacity=0] (10.72,-5.15) -- (0,0) -- (10.72,5.15) -- (7.12,0) -- cycle    ;
\draw    (525,22) -- (524.86,46.33) ;
\draw [shift={(524.84,49.33)}, rotate = 270.34000000000003] [fill={rgb, 255:red, 0; green, 0; blue, 0 }  ][line width=0.08]  [draw opacity=0] (10.72,-5.15) -- (0,0) -- (10.72,5.15) -- (7.12,0) -- cycle    ;
\draw    (525,80) -- (525.17,96.05) ;
\draw [shift={(525.2,99.05)}, rotate = 269.4] [fill={rgb, 255:red, 0; green, 0; blue, 0 }  ][line width=0.08]  [draw opacity=0] (10.72,-5.15) -- (0,0) -- (10.72,5.15) -- (7.12,0) -- cycle    ;
\draw    (525,129) -- (525.17,145.05) ;
\draw [shift={(525.2,148.05)}, rotate = 269.4] [fill={rgb, 255:red, 0; green, 0; blue, 0 }  ][line width=0.08]  [draw opacity=0] (10.72,-5.15) -- (0,0) -- (10.72,5.15) -- (7.12,0) -- cycle    ;
\draw    (525,178) -- (525.17,194.05) ;
\draw [shift={(525.2,197.05)}, rotate = 269.4] [fill={rgb, 255:red, 0; green, 0; blue, 0 }  ][line width=0.08]  [draw opacity=0] (10.72,-5.15) -- (0,0) -- (10.72,5.15) -- (7.12,0) -- cycle    ;
\draw    (525,227) -- (525.17,243.05) ;
\draw [shift={(525.2,246.05)}, rotate = 269.4] [fill={rgb, 255:red, 0; green, 0; blue, 0 }  ][line width=0.08]  [draw opacity=0] (10.72,-5.15) -- (0,0) -- (10.72,5.15) -- (7.12,0) -- cycle    ;
\draw    (525,276) -- (525.17,292.05) ;
\draw [shift={(525.2,295.05)}, rotate = 269.4] [fill={rgb, 255:red, 0; green, 0; blue, 0 }  ][line width=0.08]  [draw opacity=0] (10.72,-5.15) -- (0,0) -- (10.72,5.15) -- (7.12,0) -- cycle    ;
\draw    (525,325) -- (525.17,341.05) ;
\draw [shift={(525.2,344.05)}, rotate = 269.4] [fill={rgb, 255:red, 0; green, 0; blue, 0 }  ][line width=0.08]  [draw opacity=0] (10.72,-5.15) -- (0,0) -- (10.72,5.15) -- (7.12,0) -- cycle    ;
\draw    (525,375) -- (525.17,391.05) ;
\draw [shift={(525.2,394.05)}, rotate = 269.4] [fill={rgb, 255:red, 0; green, 0; blue, 0 }  ][line width=0.08]  [draw opacity=0] (10.72,-5.15) -- (0,0) -- (10.72,5.15) -- (7.12,0) -- cycle    ;
\draw    (525,425) -- (525.17,441.05) ;
\draw [shift={(525.2,444.05)}, rotate = 269.4] [fill={rgb, 255:red, 0; green, 0; blue, 0 }  ][line width=0.08]  [draw opacity=0] (10.72,-5.15) -- (0,0) -- (10.72,5.15) -- (7.12,0) -- cycle    ;
\draw    (525,475) -- (525.17,491.05) ;
\draw [shift={(525.2,494.05)}, rotate = 269.4] [fill={rgb, 255:red, 0; green, 0; blue, 0 }  ][line width=0.08]  [draw opacity=0] (10.72,-5.15) -- (0,0) -- (10.72,5.15) -- (7.12,0) -- cycle    ;
\draw    (526.2,30.6) -- (614,30.6) -- (614.5,560.9) -- (543.67,560.89) ;
\draw [shift={(540.67,560.89)}, rotate = 360.01] [fill={rgb, 255:red, 0; green, 0; blue, 0 }  ][line width=0.08]  [draw opacity=0] (10.72,-5.15) -- (0,0) -- (10.72,5.15) -- (7.12,0) -- cycle    ;
\draw   (508.45,560.89) .. controls (508.45,551.99) and (515.66,544.78) .. (524.56,544.78) .. controls (533.45,544.78) and (540.67,551.99) .. (540.67,560.89) .. controls (540.67,569.79) and (533.45,577) .. (524.56,577) .. controls (515.66,577) and (508.45,569.79) .. (508.45,560.89) -- cycle ;
\draw    (525,525) -- (524.86,542.33) ;
\draw [shift={(524.84,545.33)}, rotate = 270.45] [fill={rgb, 255:red, 0; green, 0; blue, 0 }  ][line width=0.08]  [draw opacity=0] (10.72,-5.15) -- (0,0) -- (10.72,5.15) -- (7.12,0) -- cycle    ;
\draw    (524.56,577) -- (524.56,593.07) ;
\draw [shift={(524.56,596.07)}, rotate = 270] [fill={rgb, 255:red, 0; green, 0; blue, 0 }  ][line width=0.08]  [draw opacity=0] (10.72,-5.15) -- (0,0) -- (10.72,5.15) -- (7.12,0) -- cycle    ;
\draw   (514,561.18) -- (535.2,561.18)(524.6,551.18) -- (524.6,571.18) ;

\draw (46,6) node [anchor=north west][inner sep=0.75pt]   [align=left] {MBConv1, k3\ding{53}3};
\draw (14,29) node [anchor=north west][inner sep=0.75pt]   [align=left] {h \ding{53} w \ding{53} c};
\draw (42,56) node [anchor=north west][inner sep=0.75pt]  [font=\large] [align=left] {DW Conv(k3\ding{53}3,c,s)};
\draw (35,107.5) node [anchor=north west][inner sep=0.75pt]  [font=\large] [align=left] {Batch Normalization};
\draw (68,160.5) node [anchor=north west][inner sep=0.75pt]  [font=\large] [align=left] {Activation};
\draw (73,212.5) node [anchor=north west][inner sep=0.75pt]  [font=\large] [align=left] {SE(r=R)};
\draw (48,265.5) node [anchor=north west][inner sep=0.75pt]  [font=\large] [align=left] {Conv(k1\ding{53}1,C,1)};
\draw (35,318.5) node [anchor=north west][inner sep=0.75pt]  [font=\large] [align=left] {Batch Normalization};
\draw (13,351.5) node [anchor=north west][inner sep=0.75pt]   [align=left] {h/s \ding{53} w/s \ding{53} C};
\draw (291,7) node [anchor=north west][inner sep=0.75pt]   [align=left] {SE(r=R)};
\draw (215,38) node [anchor=north west][inner sep=0.75pt]   [align=left] {h \ding{53} w \ding{53} c};
\draw (247,67) node [anchor=north west][inner sep=0.75pt]  [font=\large] [align=left] {Global Average Pool};
\draw (285.5,128) node [anchor=north west][inner sep=0.75pt]  [font=\large] [align=left] {Reshape};
\draw (257,186.7) node [anchor=north west][inner sep=0.75pt]   [align=left] [font=\large] {Conv(k1\ding{53}1,c/R,1)};
\draw (256,245.7) node [anchor=north west][inner sep=0.75pt]   [align=left] [font=\large] {Conv(k1\ding{53}1, c,1)};
\draw (227,100) node [anchor=north west][inner sep=0.75pt]  [font=\small] [align=left] {c};
\draw (220,158) node [anchor=north west][inner sep=0.75pt]   [align=left] {1 \ding{53} 1 \ding{53} c};
\draw (225,220) node [anchor=north west][inner sep=0.75pt]   [align=left] {1 \ding{53} 1 \ding{53} c/R};
\draw (226,340.5) node [anchor=north west][inner sep=0.75pt]   [align=left] {h \ding{53} w \ding{53} c};
\draw (476,6) node [anchor=north west][inner sep=0.75pt]   [align=left] {MBConv6, k3\ding{53}3};
\draw (436,31) node [anchor=north west][inner sep=0.75pt]   [align=left] {h \ding{53} w \ding{53} c};
\draw (469,58) node [anchor=north west][inner sep=0.75pt]  [font=\large] [align=left] {Conv(k1\ding{53}1,6c,1)};
\draw (456,109.5) node [anchor=north west][inner sep=0.75pt]  [font=\large] [align=left] {Batch Normalization};
\draw (491,156.5) node [anchor=north west][inner sep=0.75pt]  [font=\large] [align=left] {Activation};
\draw (452,203.5) node [anchor=north west][inner sep=0.75pt]  [font=\large] [align=left] {DW Conv(k3\ding{53}3,6c,s)};
\draw (456,254.5) node [anchor=north west][inner sep=0.75pt]  [font=\large] [align=left] {Batch Normalization};
\draw (491,303.5) node [anchor=north west][inner sep=0.75pt]  [font=\large] [align=left] {Activation};
\draw (476,402.5) node [anchor=north west][inner sep=0.75pt]  [font=\large] [align=left] {Conv(k1\ding{53}1,C,1)};
\draw (456,451.5) node [anchor=north west][inner sep=0.75pt]  [font=\large] [align=left] {Batch Normalization};
\draw (497,502.5) node [anchor=north west][inner sep=0.75pt]  [font=\large] [align=left] {Dropout};
\draw (430,583.5) node [anchor=north west][inner sep=0.75pt]   [align=left] {h/s \ding{53} w/s \ding{53} C};
\draw (226,286.33) node [anchor=north west][inner sep=0.75pt]   [align=left] {1 \ding{53} 1 \ding{53} c};
\draw (498,352.5) node [anchor=north west][inner sep=0.75pt]  [font=\large] [align=left] {SE(r=R)};
\draw (298.5,305.5) node [anchor=north west][inner sep=0.75pt]  [font=\Large] [align=left] {\ding{53}};

\draw (90,393) node [anchor=north west][inner sep=0.75pt]  [font=\Large] [align=left] {\large (a)};

\draw (294.5,380) node [anchor=north west][inner sep=0.75pt]  [font=\Large] [align=left] {\large (b)};

\draw (510,624) node [anchor=north west][inner sep=0.75pt]  [font=\Large] [align=left] {\large (c)};

\end{tikzpicture}

}
\caption{The basic building block of EfficientNet-B0. All MBConv blocks take the height, width, and channel of h, w, and c as input. C is the output channel of the two blocks. (Note that, MBConv= Mobile Inverted Bottleneck Convolution, DW Conv= Depth-wise Convolution, SE= Squeeze-Excitation, Conv= Convolution)
}
\label{fig:EfficientNet-B0 MBConv Block}

\end{figure}

The feature extraction of the EfficientNet-B0 baseline architecture is comprised of the several mobile inverted bottleneck convolution (MBConv)\cite {Sandler_2018_CVPR,tan_chen_pang_vasudevan_sandler_howard_le_2019} blocks with built-in squeeze-and-excitation (SE)\cite{Squeeze_and_Excitation_Networks}, Batch Normalization, and Swish activation\cite{swish_act_2017} as integrated into EfficientNet.
Compared with conventional convolution, EfficientNet's ensemble framework is, i.e., MBConv, proven to be more accurate in image classification, while reducing parameters and FLOPS by an order of magnitude.
Table\ref{EfficientNet B0 Layer Outline} shows the detailed information of each layer of the EfficientNet-B0 baseline network.
EfficientNet-B0 consists a total of of $16$ MBConv blocks varying in several aspects, for instance, kernel size, feature maps expansion phase, reduction ratio, etc.
A complete workflow of the MBConv1,k$3\times 3$ and MBConv6,k$3\times 3$ blocks are shown in Figure \ref{fig:EfficientNet-B0 MBConv Block}.
Both MBConv1,k$3\times 3$ and MBConv6,k$3\times 3$ use depthwise convolution, which integrates a kernel size of $3\times 3$ with the stride size of $s$.
In these two blocks, batch normalization, activation, and convolution with a kernel size of $1\times 1$ are integrated.
The skip connection and a dropout layer are also incorporated in MBConv6,k$3\times 3$, but this is not the case with MBConv1,k$3\times 3$. 
 Furthermore, in the case of the extended feature map, MBConv6,k$3\times 3$ is six times that of MBConv1,k$3\times 3$, and the same is true for the reduction rate in the SE block, that is, for MBConv1,k$3\times 3$ and MBConv6,k$3\times 3$, $r$ is fixed to $4$ and $24$, respectively.
Note that, MBConv6,k$5\times 5$ performs the identical operations as MBConv6,k$3\times 3$, but MBConv6,k$5\times 5$ applies a kernel size of $5\times 5$, while a kernel size of $3\times 3$ is used by MBConv6,k$3\times 3$.

\begin{table}[!ht]
\centering
\caption{EfficientNet-B0 baseline network layers outline}
\captionsetup{justification=centering}
\begin{tabular}{c|c|c|c|c} \hline
Stage      & Operator & Resolution & \#Output Feature Maps & \#Layers      \\ \hline \hline
$1$  & Conv\;$3 \times 3$    & $224 \times 224$ & $32$ & $1$ \\
$2$       & MBConv$1$,\;k$3 \times 3$    & $112 \times 112$ & $16$ & $1$ \\
$3$    & MBConv$6$,\;k$3 \times 3$    & $112 \times 112$ & $24$ & $2$\\
$4$ & MBConv$6$,\;k$5 \times 5$    & $56 \times 56$ & $40$ & $2$\\
$5$  & MBConv$6$,\;k$3 \times 3$    & $28 \times 28$ & $80$ & $3$\\
$6$       & MBConv$6$,\;k$5 \times 5$    & $14 \times 14$ & $112$ & $3$\\
$7$    & MBConv$6$,\;k$5 \times 5$    & $14 \times 15$ & $192$ & $4$\\
$8$ & MBConv$6$,\;k$3 \times 3$    & $7 \times 7$ & $320$ & $1$\\
$9$ & Conv\;$1 \times 1$ \& Pooling \& FC    & $7 \times 7$ & $1280$ & $1$\\ \hline
\end{tabular}
\label{EfficientNet B0 Layer Outline}
\end{table}

Instead of random initialization of network weights, we instantiate ImageNet's pre-trained weights in the EfficientNet model thereby accelerating the training process. 
Transferring the pre-trained weights of the ImageNet have performed a great feat in the field of image analysis, since it composes more than 14 million images covering eclectic classes.
The rationale for using pre-trained weights is that the imported model already has sufficient knowledge in the broader aspects of the image domain.
As it has been manifested in several studies \cite{Iteratively_Pruned_COVID_19,narin2020automatic}, using pre-trained ImageNet weights in the state-of-the-art CNN models remain optimistic even when the problem area (namely COVID-19 detection) is considerably distinct from the one in which the original weights have been obtained.
The optimization process will fine-tune the initial pre-training weights in the new training phase so that we can fit the pre-trained model to a specific problem domain, such as COVID-19 detection.

\subsubsection{Classifier}
\label{Classifier}

The final output of the EfficientNet architecture turns out as a global averaged feature followed by a classifier.
To perform the classification task, we used a two-layer MLP (usually called a fully connected (FC) layer), which captures the features of EfficientNet through two neural layers (each neural layer has 512 nodes).
In between FC layers, we included batch normalization, activation, and dropout layer.
Batch normalization greatly accelerates the training of deep networks and increases the stability of neural networks \cite{IoffeS15}.
It makes the optimization process smoother, resulting in a more predictable and stable gradient behavior, thereby speeding up training \cite{howdoesbatchnormalization}.
In this study, in a case of activation function, we have preferred Swish which is defined as\cite{swish_act_2017}:
\begin{equation}
\begin{aligned}
{f(x)} &= x\cdot\sigma(x)
\label{swishactivation}
\end{aligned}
\end{equation}
where $\sigma(x)=(1+exp(-x))^{-1}$ is the sigmoid function. Comparison with other activation functions Swish consistently outperforming others including Rectified Linear Unit(ReLU)\cite{conf/icml/NairH10}, which is the most successful and widely-used activation function, on deep networks applied to a variety of challenging fields i.e., image classification and machine translation.
Swish has many characteristics, such as one-sided boundedness at zero, smoothness, and non-monotonicity, which play an important role in improving it\cite{swish_act_2017}.
After performing the activation operation, we integrated a Dropout\cite{JMLR_v15_srivastava14a} layer, which is one of the preeminent regularization methods to reduce overfitting and make better predictions.
This layer can randomly drop certain FC layer nodes, which means removing all randomly selected nodes, along with all its incoming and outgoing weights.
The number of randomly selected nodes drop in each layer is obtained with a probability $p$ independent of other layers, where $p$ can be chosen by using either a validation set or a random estimate (i.e., $p=0.5$). In this study, we maintained a dropout size of $0.3$.
Next, the classification layer used the softmax activation function to render the activation from the previous FC layers into a class score to determine the class of the input chest X-ray image as COVID-19, normal, and pneumonia.
The softmax activation function is defined in the following way:
\begin{equation}
\begin{aligned}
s(y_i) &= \frac{e^{y_i}}{\sum_{j=1}^{C} e^{y_j}}
\label{softmax}
\end{aligned}
\end{equation}
where $C$ is the total number of classes. This normalization limits the output sum to $1$, so the softmax output $s(y_i)$ can be interpreted as the probability that the input belongs to the $i$ class.
In the training process, we apply the categorical cross-entropy loss function, which uses the softmax activation function in the classification layer to measure the loss between the true probability of the category and the probability of the predicted category. The categorical cross-entropy loss function is defined as
\begin{equation}
\begin{aligned}
l &= -\sum_{n=1}^{N}\log(\frac{e^{y_{i,n}}}{\sum_{j=1}^{C} e^{y_{j,n}}}).
\label{categorical cross entropy loss}
\end{aligned}
\end{equation}
The total number of input samples is denoted as $N$, and $C$ is the total number of classes, that is, $C=3$ in our case.

\subsubsection{Model Snapshots and Ensemble Prediction}
\label{Model Snapshots and Ensemble Prediction}

The main concept of building model snapshots is to train one model with constantly reducing the learning rate to attain a local minimum and save a snapshot of the current model's weight.
Later, it is necessary to actively increase the learning rate to retreat from the current local minimum requirements.
This process continues repeatedly until it completes cycles. 
One of the prominent methods for creating model snapshots for CNN is to collect multiple models during a single training run with cyclic cosine annealing \cite{huang2017snapshot}.  
The cyclic cosine annealing method starts from the initial learning rate, then gradually decreases to the minimum, and then rapidly increases.
The learning rate of cyclic cosine annealing in each epoch is defined as:
\begin{equation}
\begin{aligned}
\alpha{(t)} &= \frac{\alpha_{0}}{2}(\cos(\frac{\pi\mathrm{mod}(t-1,\ceil*{T/M})}{\ceil*{T/M}})+1)
\label{weightclassssaliencymapGradCAM}
\end{aligned}
\end{equation}
where $\alpha{(t)}$ is the learning rate at epoch $t$, $\alpha_{0}$ is the initial learning rate, $T$ is the total number of training iterations and $M$ is the number of cycles.
The weight at the bottom of each cycle is regarded as the weight of the snapshot model.
The following learning rate cycle uses these weights, but allows the learning algorithm to converge to different solutions, thereby generating diverse snapshots model.
After completing $M$ cycles of training, we get $M$ model snapshots $s_1. . .s_M$, each of which will be utilized in the ensemble prediction.

Ensemble through model snapshots is more effective than a structure based on a single model only.
Therefore, compared with the prediction of a single model, the ensemble prediction reduces the generalization error, thereby improving the prediction performance.
We have experimented with two ensemble strategies, i.e., hard ensemble and soft ensemble, to consolidate the predictions of snapshots model to classify chest X-ray images as COVID-19 or normal or pneumonia.
Both hard ensemble and soft ensemble use the last $m (m \leq M)$ model’s softmax outputs since these models have a tendency to have the lowest test error.
We also consider class weights to obtain a softmax score before applying the ensemble.
Let $O_i(x)$ be the softmax score of the test sample $x$ of the $i$-th snapshot model.
Using hard ensemble, the prediction of the $i$-th snapshot model is defined as 
\begin{equation}
\begin{aligned}
H_i &= \argmax_x O_i(x).
\label{hardvoting}
\end{aligned}
\end{equation}
The final ensemble constrains to aggregate the votes of the classification labels (i.e., COVID-19, normal, and pneumonia) in the other snapshot models and predict the category with the most votes.
On the other hand, the output of the soft ensemble includes averaging the predicted probabilities of class labels in the last $m$ snapshots model defined as
\begin{equation}
\begin{aligned}
S &= \frac{1}{m}\sum_{i=0}^{m-1} O_{M-i}(x).
\label{softvoting}
\end{aligned}
\end{equation}
Finally, the class label with the highest probability is used for the prediction.

\subsubsection{Hyper-Parameters Adjustment}
\label{Hyper-Parameters Adjustment}

Fine-tuned hyper-parameters have a great impact on the performance of the model because they directly govern the training of the model. What's more, fine-tuned parameters can avoid overfitting and form a generalized model.
Since we have dealt with an unbalanced data set, the proposed architecture may have a huge possibility to confront the problem of overfitting.
In order to solve the problem of overfitting, we use $L1L2$ weight decay regularization with coefficients $1e-5$ and $1e-3$ in FC layers.
Next, dropout is another successful regularization technique that has been integrated into the proposed architecture, especially in FC layers with $p = 0.3$, to suppress overfitting.
In the experiments on the proposed architecture, we have explored the Adam \cite{kingma2014adam}  optimizer, which can converge faster.
When creating snapshots, we set the number of epochs to $25$, the minimum batch size to $8$, the initial learning rate to $1e-4$, and the number of cycles to $5$, thus providing $5$ snapshots for each model, on which we build up the ensemble prediction.

\subsubsection{Visual Explanations using Grad-CAM}
\label{Visualizations}

Although the CNN-based modular architecture provides encouraging recognition performance for image classification, there are still several issues where it is challenging to reveal why and how to produce such impressive results.
Due to its black-box nature, it is sometimes contrary to apply it in a medical diagnosis system where we need an interpretable system i.e., visualization as well as an accurate diagnosis.
Despite it has certain challenges, researchers are still endeavoring to seek for an efficient visualization technique since it can contribute the most critical key facts in the health-care system into focus, assist medical practitioners to distinguish correlations and patterns in imaging, and perform data analysis more efficacious.
In the field of detecting COVID-19 through chest X-rays, some early studies focused on visualizing the behavior of CNN models to distinguish between different categories (such as COVID-19, normal, and pneumonia), so they can produce explanatory models.
In our proposed model, we applied a gradient-based approach named Grad-CAM\cite{8237336}, which measures the gradients of features maps in the final convolution layer on a CNN model for a target image, to foreground the critical regions that are class-discriminative saliency maps.
In Grad-CAM, gradients that are flowing back to the final convolutional layer in a CNN model are globally averaged to calculate the target class weights of each filter.
Grad-CAM heat-map is a combination of weighted feature maps, followed by a ReLU activation.
The class-discriminative saliency map $L^{c}$ for the target image class $c$ is defined as follows\cite{8237336}:
\begin{equation}
\begin{aligned}
L^{c}_{i,j} &= ReLU(\sum_{k}w^{c}_{k}A^{k}_{i,j}),
\label{classssaliencymapGradCAM}
\end{aligned}
\end{equation}
where $A^{k}_{i,j}$ denotes the activation map for the $k$-th filter at a spatial location ($i,j$), and ReLU captures the positive features of the target class.
The target class weights of $k$-th filter is computed as:
\begin{equation}
\begin{aligned}
w^{c}_{k} &= \frac{1}{Z}\sum_{i}\sum_{j}\frac{\partial Y^{c}}{\partial A^{k}_{i,j}},
\label{weightclassssaliencymapGCAM}
\end{aligned}
\end{equation}
where $Y^{c}$ is the probability of classifying the target category as $c$, and the total number of pixels in the activation map is denoted as $Z$.

\section{Experiments and Results}
\label{Experiments and Results}

In this section, we will present the results and consider several experimental settings to analyze the results of the proposed ECOVNet to explore the robustness of the model.
The performance of the proposed model for figuring out the three-class classification problem is compared with some state-of-the-art methods. 
The three-class classification problem is to determine whether the chest X-ray image belongs to the category of COVID-19 or the normal or pneumonia category.
All our programs are written in Python, and the software pile is composed of Keras with the TensorFlow backend and scikit-learn.


\subsection{Data Set Settings}
\label{Data Set Settings}

In sub-section \ref{Dataset} and sub-section \ref{Data Augmentation}, the benchmark data set with the augmentation approach used in the experiment is illustrated in brief.
We configure two test sets (namely Imbalanced and Balanced Test) that the imbalanced test is the original test set that comes from COVIDx while the balanced test is also from COVIDx test set, but we randomly choose $100$ images for both normal and pneumonia where the test size of COVID-19 is fixed, i.e., $100$.
During training, we set the training and validation ratios to 90\% and 10\%, respectively.
The entire image distribution of training, validation, and testing is shown in Table \ref{image partition balance and imbalance test}.
We regard the pre-trained EfficientNet as feature extraction, and in the description of the related structure in sub-section \ref{Proposed Network Architecture}, the impression is that EfficientNet is a series of models formed by arbitrary selection of scale factors.
In our experiment, we consider EfficientNet B0 to B5 base models; however, the input shapes are different.
Table \ref{EfficientNet Configuration} displays a list of input shapes for each base model as well as the total number of parameters during training.

\begin{table}[!ht]
\small
\centering
\caption{Image partition of Training, Validation, and Testing set for Balanced and Imbalanced test}
\captionsetup{justification=centering}
      \begin{tabular}{c|ccc|c}
        \hline
        Category  & COVID-19  & Normal   & Pneumonia & Total\\ \hline \hline
        Training   & $441$        & $7,170$        & $4,914$    & $12,525$ \\ \hline
        Validation   & $48$        & $796$        & $545$                & $1,389$ \\ \hline
        Testing(Balanced) & $100$        & $100$        & $100$                 & $300$ \\ \hline
        Testing(Imbalanced) & $100$        & $885$        & $594$                 & $1,579$ \\ \hline
      \end{tabular}
      \label{image partition balance and imbalance test}
\end{table}

\begin{table}[!ht]
\small
\centering
\caption{Image resolution and total number of parameters of ECOVNet considering the base models (B0 to B5) of EfficientNet}
\captionsetup{justification=centering}
      \begin{tabular}{ccc||ccc}
        \hline
        Base Model  & Image Resolution  & \makecell{Parameter Size \\(ECOVNet)}  & Base Model  & Image Resolution  & \makecell{Parameter Size \\(ECOVNet)}\\ \hline \hline
        EfficientNet-B0  & $224\times 224$  & $4,978,847$   & EfficientNet-B3  & $360\times 360$  & $11,844,907$ \\ \hline
        EfficientNet-B1  & $240\times 240$  & $7,504,515$   & EfficientNet-B4  & $380\times 380$  & $18,867,291$ \\ \hline
        EfficientNet-B2  & $260\times 260$  & $8,763,893$   & EfficientNet-B5  & $456\times 456$  & $29,839,091$ \\ \hline
      \end{tabular}
      \label{EfficientNet Configuration}
\end{table}

\subsection{Evaluation Metrics}
\label{Evaluation Metrics}
In order to evaluate the performance of the proposed method, we considered the following evaluation metrics: accuracy, precision, recall, F1 score, confidence interval (CI), receiver operating characteristic (ROC) curve and area under the curve (AUC). The definitions of accuracy, precision, recall and F1 score are as follows:
\begin{equation}
\begin{aligned}
Accuracy &= \frac{TP+TN}{Total\,Samples}
\label{accuracy}
\end{aligned}
\end{equation}
\begin{equation}
\begin{aligned}
Precision &= \frac{TP}{TP+FP}
\label{Precision}
\end{aligned}
\end{equation}
\begin{equation}
\begin{aligned}
Recall &= \frac{TP}{TP+FN}
\label{Recall}
\end{aligned}
\end{equation}
\begin{equation}
\begin{aligned}
F1 &= 2\times\frac{Precision \times Recall }{Precision + Recall }
\label{F1}
\end{aligned}
\end{equation}

where $TP$ stands for true positive, while $TN$, $FP$, and $FN$ stand for true negative, false positive, and false negative, respectively. 
Since the benchmark data set is not balanced, $F1$ score may be a more substantial evaluation metric. For example, COVID-19 has $589$ images and non-COVID, that is, normal and pneumonia have $8,851$ and $6,053$ images, respectively.
What's more, a $95\%$ CI is considered as it's a more practical metric compared with specific performance indicators. It can increase the level of statistical significance as well as can reflect the reliability of the problem domain.
Finally, we displayed the ROC curve to display the results and measured the area under the ROC curve (usually called AUC) to provide information about the effectiveness of the model.
The ROC curve is plotted between True Positive Rate (TPR)/Recall and False Positive Rate (FPR), and FPR is defined as

\begin{equation}
\begin{aligned}
FPR &= \frac{FP}{FP+TN}.
\label{False Positive Rate}
\end{aligned}
\end{equation}

\subsection{Prediction performance of proposed ECOVNet}
\label{Prediction performance of proposed ECOVNet}

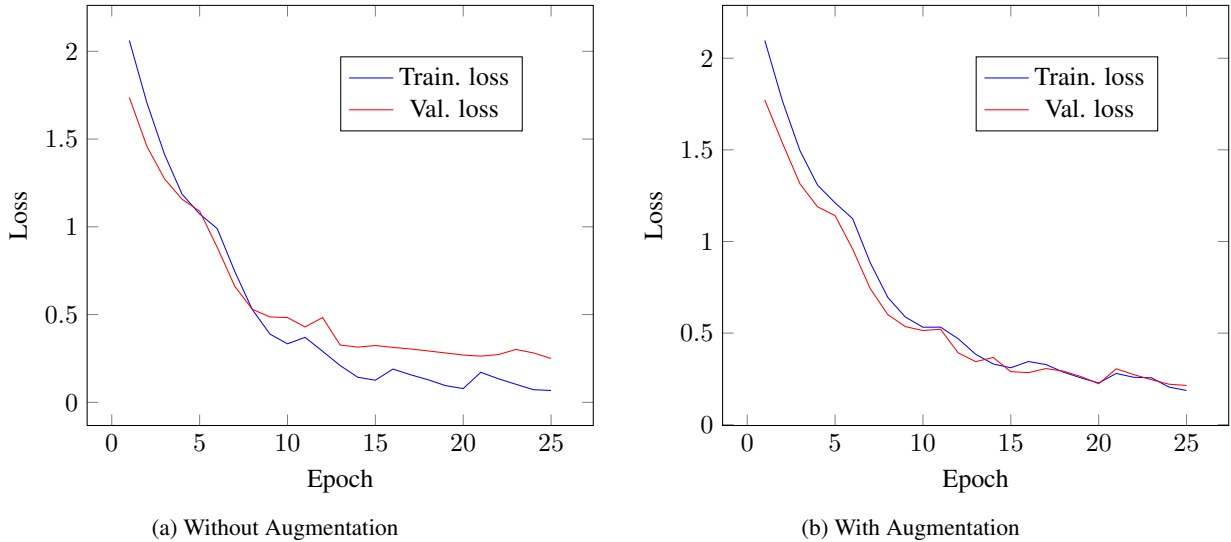
\begin{figure}[!ht]
\centering
\resizebox{16cm}{!}{

\pgfplotsset{
legend style={at={(0.5,0.7)},
anchor=south west
}}

\begin{subfigure}{0.45\linewidth}

\begin{tikzpicture}
\begin{axis}[xlabel=Epoch,ylabel=Loss]

\addplot[blue] coordinates {

(1,2.062086343765259)
(2,1.7064379453659058)
(3,1.4134962558746338)
(4,1.1860666275024414)
(5,1.0734111070632935)
(6,0.9900187253952026)
(7,0.7454519867897034)
(8,0.5285671353340149)
(9,0.38884106278419495)
(10,0.3335166871547699)
(11,0.37027281522750854)
(12,0.29072949290275574)
(13,0.20927345752716064)
(14,0.14300593733787537)
(15,0.12603759765625)
(16,0.18972700834274292)
(17,0.15734925866127014)
(18,0.12866967916488647)
(19,0.09512544423341751)
(20,0.07843358814716339)
(21,0.17104333639144897)
(22,0.13444212079048157)
(23,0.10282312333583832)
(24,0.07200257480144501)
(25,0.06748250871896744)

};

\addplot[red] coordinates {

(1,1.7369626760482788)
(2,1.4585888385772705)
(3,1.2735384702682495)
(4,1.157288670539856)
(5,1.0893895626068115)
(6,0.8834366202354431)
(7,0.6613011360168457)
(8,0.5300467610359192)
(9,0.48647379875183105)
(10,0.4830965995788574)
(11,0.42943504452705383)
(12,0.4830395579338074)
(13,0.32635244727134705)
(14,0.31473806500434875)
(15,0.3236863911151886)
(16,0.3129735589027405)
(17,0.30412033200263977)
(18,0.2929728925228119)
(19,0.2812096178531647)
(20,0.26955223083496094)
(21,0.2633296251296997)
(22,0.27211877703666687)
(23,0.30133670568466187)
(24,0.2813869118690491)
(25,0.24981658160686493)

};

\legend{ Train. loss,Val. loss}
\end{axis}
\end{tikzpicture}
\caption{Without Augmentation}
\end{subfigure}

\hspace{1cm}

\begin{subfigure}{0.45\linewidth}
\begin{tikzpicture}
\begin{axis}[xlabel=Epoch,ylabel=Loss]

\addplot[blue] coordinates {

(1,2.0972774028778076)
(2,1.7679870128631592)
(3,1.4960417747497559)
(4,1.3079050779342651)
(5,1.2108970880508423)
(6,1.1246377229690552)
(7,0.8841667175292969)
(8,0.6948858499526978)
(9,0.5877425074577332)
(10,0.5321572422981262)
(11,0.532623827457428)
(12,0.4697010815143585)
(13,0.384623646736145)
(14,0.33168354630470276)
(15,0.3107251822948456)
(16,0.34495407342910767)
(17,0.32836389541625977)
(18,0.28693273663520813)
(19,0.25543761253356934)
(20,0.2279692441225052)
(21,0.27981138229370117)
(22,0.2588455080986023)
(23,0.25604239106178284)
(24,0.2054705172777176)
(25,0.18685218691825867)

};

\addplot[red] coordinates {

(1,1.772192358970642)
(2,1.5382708311080933)
(3,1.3147636651992798)
(4,1.1894134283065796)
(5,1.1412540674209595)
(6,0.9583495259284973)
(7,0.743082582950592)
(8,0.6008408069610596)
(9,0.5355625748634338)
(10,0.5140117406845093)
(11,0.5211459398269653)
(12,0.3925231993198395)
(13,0.34438982605934143)
(14,0.36685821413993835)
(15,0.2897740602493286)
(16,0.2841746509075165)
(17,0.3065350651741028)
(18,0.29215556383132935)
(19,0.26278209686279297)
(20,0.22357617318630219)
(21,0.3052757978439331)
(22,0.2734275460243225)
(23,0.2458520084619522)
(24,0.22122733294963837)
(25,0.214264377951622)

};

\legend{ Train. loss,Val. loss}
\end{axis}
\end{tikzpicture}
\caption{With Augmentation}
\end{subfigure}
}

\caption{Loss curve of ECOVNet (Base model EfficientNet-B5) during training}
\label{Loss curve of EfficientNet-B5}

\end{figure}

In Table \ref{ECOVNet without using ensemble}, the predictions of the proposed ECOVNet without any ensemble are shown.
In the comparison without an ensemble, the prediction of ECOVNet with EfficientNet-B5 pre-trained weights yields superior results than other base models for the case of images with augmentation and without augmentation, which reflects the fact that feature extraction using an optimized model that considers three aspects, namely higher depth and width, and a broader image resolution, can capture more and finer details, thereby improving classification accuracy.
Without augmentation, under the condition of the imbalanced test set, ECOVNet's accuracy reaches $96.26\%$, and its performance is slightly better for the balance test set, reaching $96.33\%$ accuracy.
On the other hand, under augmentation condition, ECOVNet has the same best accuracy, i.e., $94.68\%$ for both unbalanced and balanced test sets.
Moreover, in Table \ref{ECOVNet without using ensemble}, we used a $95\%$ CI for accuracy as the measure to analyze the uncertainty inherent of the ECOVNet.
A tight range of CI means higher precision, while the wide range of CI indicates the opposite.
As we can see, for the imbalanced test set, the CI interval is within a narrow range, but for the balanced case, the CI range is wider because it considers a smaller amount of test data.
Furthermore, Figure \ref{Loss curve of EfficientNet-B5} shows the training loss of ECOVNet considering EfficientNet-B5.
Note that the value in \textbf{bold} indicates that the method has statistically better performance than other methods.

We implement two ensemble strategies: hard ensemble and soft ensemble, and each ensemble considers a total of $5$ model snapshots that are generated during a single training. 
Table \ref{Class-wise results of EfficientNet-B5 wo augmentation} and Table \ref{Class-wise results of EfficientNet-B5 w augmentation} show the classification results of different evaluation indicators for without augmentation and augmentation, respectively, including ensemble methods and no ensemble.
As shown in Table \ref{Class-wise results of EfficientNet-B5 wo augmentation}, in handling COVID-19 cases, the ensemble methods are significantly better than the no ensemble method.
More specifically, the recall hits its maximum value, which is $100\%$, and to a greater extent, this result demonstrates the robustness of our proposed architecture.
In addition, considering that the test set is balanced, soft integration appears to be the preferred method because of its precision, recall, and F1 score of $100\%$.
Comparing two ensembles, since the average softmax score of each category will affect the direction of the desired result, the effect of the soft ensemble is better than the hard ensemble.
Owing to the uneven distribution of the imbalanced test set, an F1 score may be more reliable than an accuracy.
It can be clearly seen from Table \ref{Class-wise results of EfficientNet-B5 wo augmentation} that for the unbalanced test set, compared with no ensemble, the ensemble methods can improve the F1 score of COVID-19, while the F1 scores of the hard ensemble and soft ensemble are $95.57\%$ and $96.15\%$, respectively.
For augmentation, in Table \ref{Class-wise results of EfficientNet-B5 w augmentation}, we see that the ensemble method presents better results than the no ensemble, leading to the exception that the hard ensemble is slightly better than the soft ensemble.
However, for augmentation and without augmentation, with an imbalanced test set, we observe that accuracy with more precision than a balanced test set, so the confidence interval is tight when computed from an imbalanced test set since it covers a large sample.

\begin{table}[!ht]
\small
\begin{minipage}{\textwidth}
\centering
\caption{Prediction performance of proposed ECOVNet without using ensemble}
\begin{tabular}{cc||cccc||cccc}
\hline
\multicolumn{2}{c}{}                        & \multicolumn{4}{c}{Imbalance Test} & \multicolumn{4}{c}{Balance Test}  \\ \hline \hline
Method                   & Pre-trained Weight & \makecell{Precision\\($\%$)}  & \makecell{Recall\\($\%$)} & \makecell{F1\\($\%$)} & \makecell{Accuracy\\($\%$)($95\%$ CI)} & \makecell{Precision\\($\%$)} & \makecell{Recall\\($\%$)} & \makecell{F1\\($\%$)} & \makecell{Accuracy\\($\%$)($95\%$ CI)} \\ \hline
\multirow{6}{*}{\makecell{ECOVNet\\(w/o aug.\footnote{w/o aug.=without augmentation})}} & EfficientNet-B0    &   $93.27$	        & $93.29$       &  	$93.27$   &  $93.29\pm 1.23$        &  $89.11$		         &  $88.33$      &  $88.41$  & $88.33\pm 3.63$         \\ \cline{2-10} 
                         & EfficientNet-B1    &   $94.28$	        &$94.30$      &$94.26$     &     $94.30\pm 1.14$     &  $91.20$		         & $90.33$       & $90.27$   &  $90.33\pm 3.34$       \\ \cline{2-10} 
                         & EfficientNet-B2    &  $93.24$          & $93.03$       & $93.08$   &  $93.03\pm 1.26$        &$91.87$           & $91.67$       & $91.70$   &   $91.67\pm 3.13$       \\ \cline{2-10} 
                         & EfficientNet-B3    &  $95.56$          & $95.57$       &  $95.56$  &    $95.57\pm 1.01$      &    $94.94$       & $94.67$       & $94.65$   &  $94.67\pm 2.54$        \\ \cline{2-10} 
                         & EfficientNet-B4    & $95.52$         &$95.50$        &$95.50$    & $95.50\pm 1.02$         & $95.26$          &$95.00$       &$95.01$   & $95.00\pm 2.47$         \\ \cline{2-10} 
                             & EfficientNet-B5    &   $\textbf{96.28}$         &    $\textbf{96.26}$    & $\textbf{96.26}$    & $\textbf{96.26}\pm 0.94$          &   $\textbf{96.44}$        & $\textbf{96.33}$       &  $\textbf{96.34}$  &   $\textbf{96.33}\pm 2.13$       \\ \hline
\multirow{6}{*}{\makecell{ECOVNet\\(w/ aug.\footnote{w/ aug.=with augmentation})}} & EfficientNet-B0  &  $91.71$          & $74.10$       & $79.72$   &  $74.10\pm 2.16$        &   $85.59$        &$80.67$        & $80.87$   &  $80.67\pm 4.47$   \\ \cline{2-10} 
                         & EfficientNet-B1    &  $91.02$          & $86.19$       & $87.67$   &  $86.19\pm 1.70$        &$89.24$     & $88.67$   &  $88.64$ &$88.67\pm 3.59$       \\ \cline{2-10} 
                         & EfficientNet-B2    & $93.60$           &$93.10$        & $93.24$   & $93.10\pm 1.25$         &     $91.80$      & $91.67$       & $91.66$   &    $91.67\pm 3.13$      \\ \cline{2-10} 
                         & EfficientNet-B3    &    $92.60$        &   $90.25$     & $90.92$   &     $90.25\pm 1.46$     &   $91.73$        &$91.67$        & $91.64$   &    $91.67\pm 3.13$      \\ \cline{2-10} 
                         & EfficientNet-B4    &   $94.32$         &   $93.73$     & $93.89$   &     $93.73\pm 1.20$     &   $94.01$        & $94.00$       & $94.00$   &    $94.00\pm 2.69$      \\ \cline{2-10} 
                         & EfficientNet-B5    &      $94.79$      &  $94.68$      &$94.70$    &  $94.68\pm 1.11$        &  $94.82$         & $94.67$       & $94.68$   &  $94.67\pm 2.54$        \\ \hline
\end{tabular}
\label{ECOVNet without using ensemble}
\end{minipage}
\end{table}

\begin{table}[!ht]
\small
\begin{minipage}{\textwidth}
\centering
\caption{Class-wise classification results of ECOVNet (Base model EfficientNet-B5) without augmentation}
\begin{tabular}{cc||cccc||cccc}
\hline
\multicolumn{2}{c}{}               & \multicolumn{4}{c}{Imbalance Test}         & \multicolumn{4}{c}{Balance Test}           \\ \hline \hline
Method                   & Class     & \makecell{Precision\\($\%$)} & \makecell{Recall\\($\%$)} & \makecell{F1\\($\%$)} & \makecell{Accuracy\\($\%$)($95\%$ CI)}          & \makecell{Precision\\($\%$)} & \makecell{Recall\\($\%$)} & \makecell{F1\\($\%$)} & \makecell{Accuracy\\($\%$)($95\%$ CI)}          \\ \hline
\multirow{3}{*}{\makecell{ECOVNet\\(W/O Ensemble)}} & COVID-19  &   $91.43$        &$96.00$        &  $93.66$  & \multirow{3}{*}{$96.26\pm 0.94$} &    $\textbf{100}$       & $96.00$       &$97.96$    & \multirow{3}{*}{$96.33\pm 2.13$} \\ \cline{2-5} \cline{7-9}
                         & Normal    &  $97.07$         &$97.29$        & $97.18$   &                   &   $93.40$ &$99.00$          & $96.12$       &                   \\ \cline{2-5} \cline{7-9}
                         & Pneumonia &   $95.91$        & $94.78$       & $95.34$   &                   &  $95.92$  &$94.00$           & $94.95$       &                  \\ \hline
\multirow{3}{*}{\makecell{ECOVNet\\(Hard Ensemble)}} & COVID-19  &$94.17$           & $97.00$       &$95.57$    & \multirow{3}{*}{$96.07\pm 0.96$} &  $\textbf{100}$         & $97.00$       &$98.48$    & \multirow{3}{*}{$95.67\pm 2.30$} \\ \cline{2-5} \cline{7-9}
                         & Normal    &$97.05$           & $96.72$       &$96.89$    &                   &     $93.27$      &$97.00$        & $95.10$   &                   \\ \cline{2-5} \cline{7-9}
                         & Pneumonia & $94.95$          & $94.95$       &$94.95$    &                   & $93.94$          &$93.00$        & $93.47$   &                   \\ \hline
\multirow{3}{*}{\makecell{ECOVNet\\(Soft Ensemble)}} & COVID-19  & $92.59$          & $\textbf{100}$       & $96.15$   & \multirow{3}{*}{$96.07\pm 0.96$} & $\textbf{100}$          & $\textbf{100}$       & $\textbf{100}$   & \multirow{3}{*}{$\textbf{97.00}\pm 1.93$} \\ \cline{2-5} \cline{7-9}
                         & Normal    &$97.05$           & $96.61$       & $96.83$   &                   & $93.33$          & $98.00$       & $95.61$   &                   \\ \cline{2-5} \cline{7-9}
                         & Pneumonia & $95.25$          & $94.61$       & $94.93$   &                   &   $97.89$        & $93.00$       & $95.38$   &                   \\ \hline
\end{tabular}
\label{Class-wise results of EfficientNet-B5 wo augmentation}
\end{minipage}

\end{table}

\begin{table}[!ht]
\small
\begin{minipage}{\textwidth}
\centering
\caption{Class-wise classification results of ECOVNet (Base model EfficientNet-B5) with augmentation}
\begin{tabular}{cc||cccc||cccc}
\hline
\multicolumn{2}{c}{}               & \multicolumn{4}{c}{Imbalance Test}         & \multicolumn{4}{c}{Balance Test}           \\ \hline \hline
Method                   & Class     & \makecell{Precision\\($\%$)} & \makecell{Recall\\($\%$)} & \makecell{F1\\($\%$)} & \makecell{Accuracy\\($\%$)($95\%$ CI)}          & \makecell{Precision\\($\%$)} & \makecell{Recall\\($\%$)} & \makecell{F1\\($\%$)} & \makecell{Accuracy\\($\%$)($95\%$ CI)}          \\ \hline
\multirow{3}{*}{\makecell{ECOVNet\\(W/O Ensemble)}} & COVID-19  &  $87.62$         &   $92.00$     & $89.76$   & \multirow{3}{*}{$94.68\pm 1.11$} & $98.92$          &$92.00$        & $95.34$   & \multirow{3}{*}{$94.67\pm 2.54$} \\ \cline{2-5} \cline{7-9}
                         & Normal    & $97.31$          &  $94.12$      &  $95.69$  &                   & $91.35$          &  $95.00$      & $93.14$   &                   \\ \cline{2-5} \cline{7-9}
                         & Pneumonia &  $97.31$         & $95.96$       & $94.06$   &                   &$94.17$           & $\textbf{97.00}$       &$95.57$   &                   \\ \hline
\multirow{3}{*}{\makecell{ECOVNet\\(Hard Ensemble)}} & COVID-19  &$90.29$           & $93.00$       &$91.63$    & \multirow{3}{*}{$\textbf{95.50}\pm 1.02$} &$98.94$           &$93.00$        &  $95.88$  & \multirow{3}{*}{$95.33\pm 2.39$} \\ \cline{2-5} \cline{7-9}
                         & Normal    &  $97.35$         &$95.37$        & $96.35$   &                   & $90.65$          &$\textbf{97.00}$        &$93.72$ &                   \\ \cline{2-5} \cline{7-9}
                         & Pneumonia & $93.76$          &$96.13$        & $94.93$   &                   & $96.97$          & $96.00$       & $\textbf{96.48}$   &                   \\ \hline
\multirow{3}{*}{\makecell{ECOVNet\\(Soft Ensemble)}} & COVID-19  &  $85.45$         & $94.00$       & $89.52$   & \multirow{3}{*}{$95.19\pm 1.06$} &  $\textbf{98.95}$         & $94.00$       &$96.41$    & \multirow{3}{*}{$95.00\pm 2.47$} \\ \cline{2-5} \cline{7-9}
                         & Normal    & $97.67$          & $94.92$       & $96.28$   &                   &  $92.23$         &      $95.00$  &$93.60$    &                   \\ \cline{2-5} \cline{7-9}
                         & Pneumonia &  $93.43$         & $95.79$       & $94.60$   &                   & $94.12$          &     $96.00$   & $95.05$   &                   \\ \hline
\end{tabular}
\label{Class-wise results of EfficientNet-B5 w augmentation}
\end{minipage}
\end{table}

It can be seen from Figure \ref{Comparison ensemble and no ensemble ECOVNet accuracy} that since the ensemble methods combine the predictions from the model snapshots, the ensemble methods tend to improve the classification accuracy of the proposed ECOVNet.
In addition, it is obvious that when we consider deeper base models, the classification accuracy of the proposed ECOVNet will increase. 
More specifically, in the case of a soft ensemble, the base models EfficientNet-B4 and EfficientNet-B5 provide the same accuracy and have the highest accuracy, that is, $97\%$.
Meanwhile, when the base model is moderately deeper, the hard ensemble and the soft ensemble have comparable results. On the other hand, when the model is deeper, the soft ensemble shows its superiority.
Taking into account COVID-19 cases, for the balanced test data with soft ensemble, Figure \ref{Precision Recall F1 ECOVNet balanced soft ensemble COVID-19} shows the precision, recall, and F1 score of ECOVNet.
When comparing the precision of ECOVNet, we have seen that, except for EfficientNet-B0, almost all base models show significantly better performance.
However, in terms of recall, as we consider more in-depth base models, the value gradually increases, but it decreases by $4\%$ from ECOVNet-B0 to ECOVNet-B1.
The same observation is true for F1-score while a drop of $1\%$ from ECOVNet-B0 to ECOVNet-B1.

\definecolor{C3}{RGB}{145,240,198}
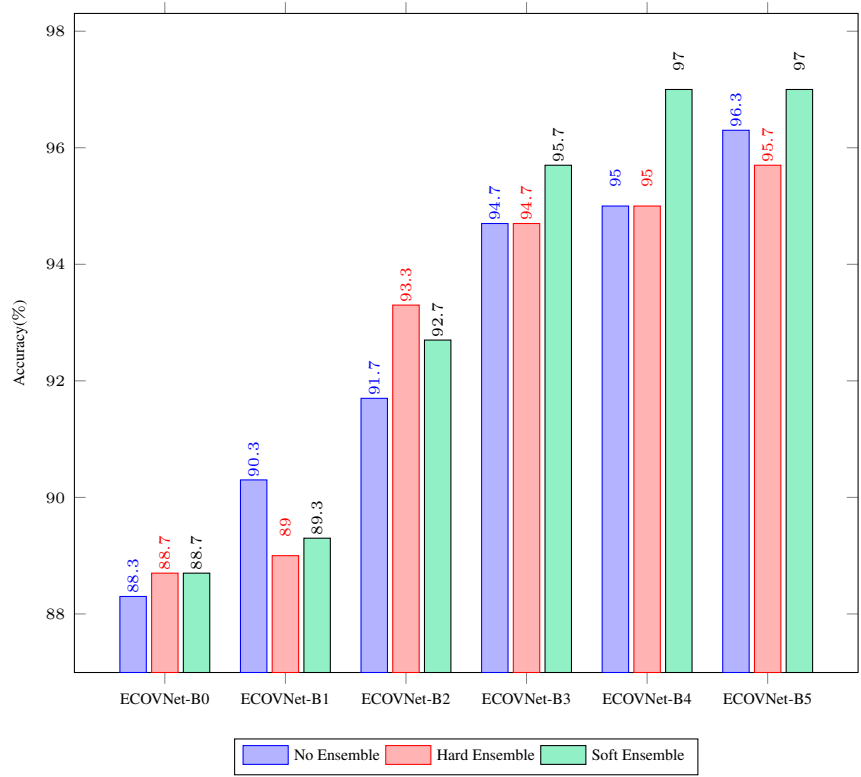
\begin{figure}[htbp]
\centering
\begin{tikzpicture}
[scale=1]
\begin{axis}[ybar,
    legend image code/.code={
    \draw [#1] (0cm,-0.1cm) rectangle (0.6cm,0.15cm); },
    enlargelimits=0.15,
    legend style={at={(0.5,-0.10)},
      anchor=north,legend columns=-1},
    ylabel={Accuracy($\%$)},font=\footnotesize,
    symbolic x coords={ECOVNet-B0,ECOVNet-B1,ECOVNet-B2,ECOVNet-B3,ECOVNet-B4,ECOVNet-B5},font=\tiny,
    xtick=data,
    nodes near coords,
    nodes near coords align={vertical},
    nodes near coords style={font=\tiny,yshift=4ex},
    width=12cm,
    every node near coord/.append style={
                anchor=east,
                rotate=90
        }
    ]
    \addplot
	coordinates {(ECOVNet-B0,88.3) (ECOVNet-B1,90.3)
		 (ECOVNet-B2,91.7) (ECOVNet-B3,94.7) (ECOVNet-B4,95.00) (ECOVNet-B5,96.3)};
\addplot
	coordinates {(ECOVNet-B0,88.7) (ECOVNet-B1,89.00) 
		(ECOVNet-B2,93.3) (ECOVNet-B3,94.7) (ECOVNet-B4,95.00) (ECOVNet-B5,95.7)};
\addplot[fill=C3] 
	coordinates {(ECOVNet-B0,88.7) (ECOVNet-B1,89.3) 
		(ECOVNet-B2,92.7) (ECOVNet-B3,95.7) (ECOVNet-B4,97.00) (ECOVNet-B5,97.00)};
\legend{No Ensemble, Hard Ensemble, Soft Ensemble}
\end{axis}
\end{tikzpicture}
\caption{Comparison between ensemble and no ensemble of the proposed ECOVNet in terms of accuracy for the balanced test data.}
\label{Comparison ensemble and no ensemble ECOVNet accuracy}
\end{figure}

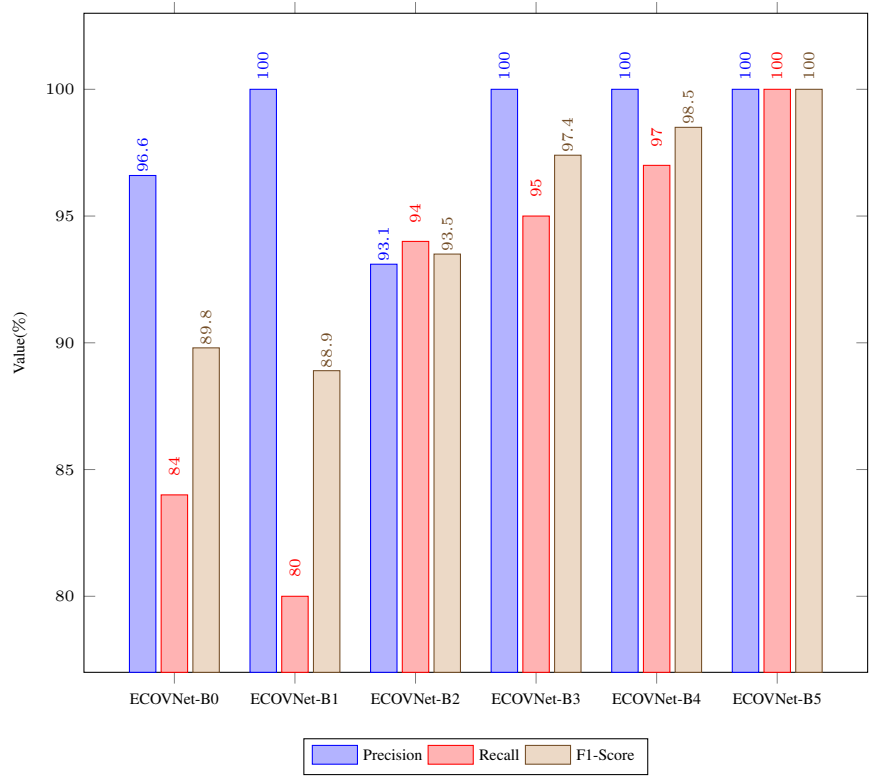
\begin{figure}[htbp]
\centering
\begin{tikzpicture}
[scale=1]
  \begin{axis}[
    ybar,
    legend image code/.code={
    \draw [#1] (0cm,-0.1cm) rectangle (0.6cm,0.15cm); },
    enlargelimits=0.15,
    legend style={at={(0.5,-0.10)},
      anchor=north,legend columns=-1},
    ylabel={Value($\%$)},font=\footnotesize,
    symbolic x coords={ECOVNet-B0,ECOVNet-B1,ECOVNet-B2,ECOVNet-B3,ECOVNet-B4,ECOVNet-B5},font=\tiny,
    xtick=data,
    nodes near coords,
    nodes near coords align={vertical},
    nodes near coords style={font=\tiny,yshift=4ex},
    width=12cm,
    every node near coord/.append style={
                anchor=east,
                rotate=90
        }
    ]
    \addplot
	coordinates {(ECOVNet-B0,96.6) (ECOVNet-B1,100)
		 (ECOVNet-B2,93.1) (ECOVNet-B3,100) (ECOVNet-B4,100) (ECOVNet-B5,100)};
	\addplot 
	coordinates {(ECOVNet-B0,84) (ECOVNet-B1,80)
		 (ECOVNet-B2,94) (ECOVNet-B3,95) (ECOVNet-B4,97) (ECOVNet-B5,100)};
	\addplot 
	coordinates {(ECOVNet-B0,89.8) (ECOVNet-B1,88.9)
		 (ECOVNet-B2,93.5) (ECOVNet-B3,97.4) (ECOVNet-B4,98.5) (ECOVNet-B5,100)};
\legend{Precision,Recall,F1-Score}
  \end{axis}
\end{tikzpicture}
\caption{Precision, Recall, F1 score of the proposed ECOVNet for the balanced test data with soft ensemble considering COVID-19 cases}
\label{Precision Recall F1 ECOVNet balanced soft ensemble COVID-19}
\end{figure}

\begin{figure}[htbp]
\centering
\includegraphics[width=1\textwidth]{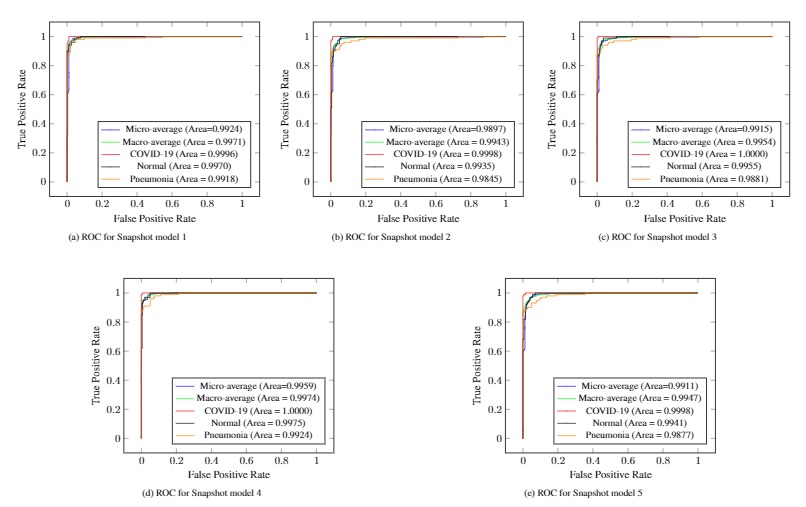}
\caption{ROC curves of model snapshots of the proposed ECOVNet considering EfficientNet-B5 base model}
\label{ROC curve for ECOVNet soft balance}
\end{figure}

\begin{figure}[htbp]
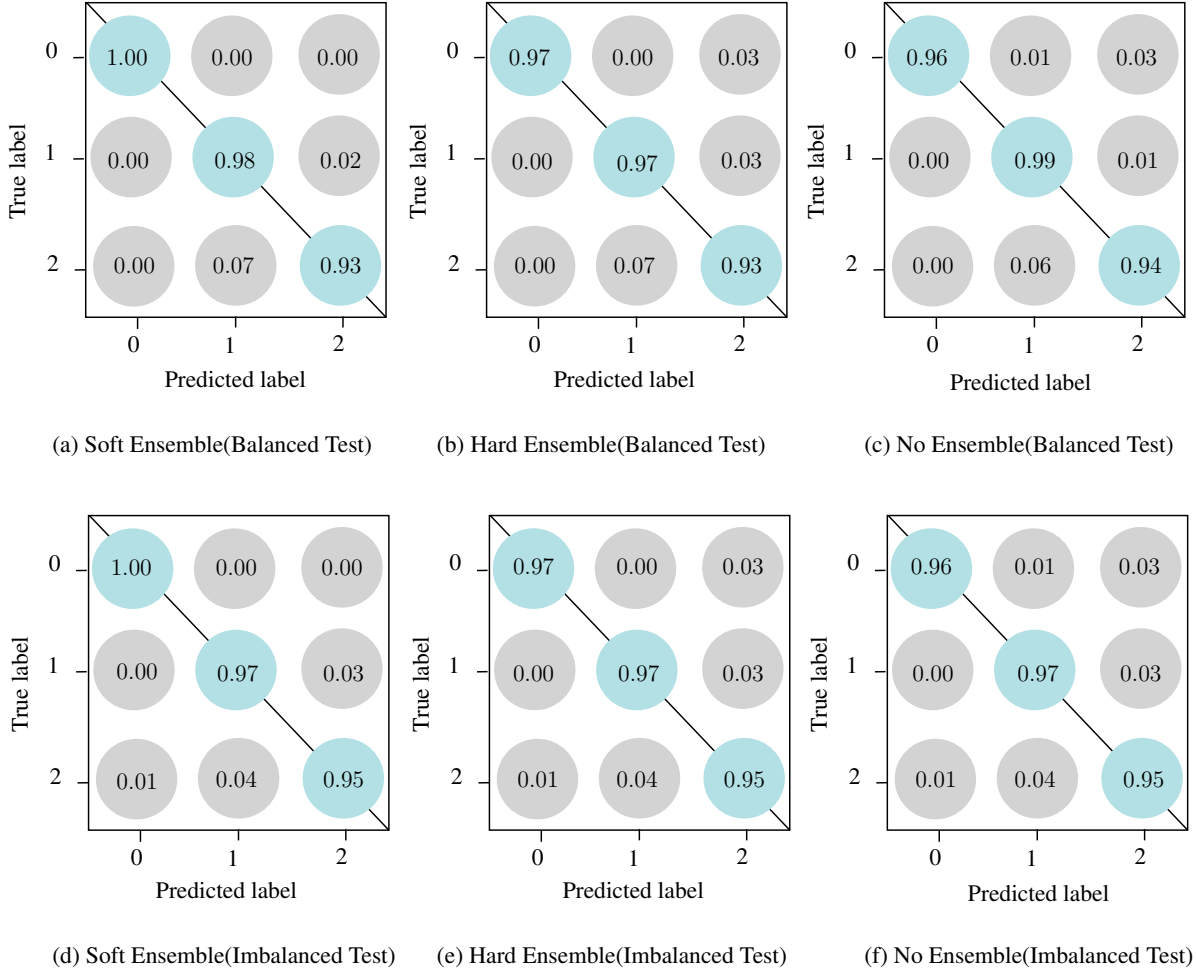

\centering
\resizebox{16cm}{!}{

\tikzset{every picture/.style={line width=0.75pt}} 



}
\caption{Confusion matrices of the proposed ECOVNet considering EfficientNet-B5 as a base model. In the confusion matrices, the predicted labels, such as COVID-19, Normal, and Pneumonia, are marked as 0, 1, and 2, respectively.}

\label{Confusion matrices ECOVNet EfficientNet-B5}

\end{figure}

It is often useful to analyze the ROC curve to reflect the classification performance of the model since the ROC curve gives a summary of the trade-off between the true positive rate and the false positive rate of a model that takes into account different probability thresholds.
In Figure \ref{ROC curve for ECOVNet soft balance}, the ROC curves show the micro and macro average and class-wise AUC scores obtained by the proposed ECOVNet, where each curve refers to the ROC curve of an individual model snapshot.
The AUC scores of all categories are consistent, indicating that the prediction of the proposed model is stable. However, the AUC scores in the third  and fourth snapshots are better than other snapshots.
As it is evident from Figure \ref{ROC curve for ECOVNet soft balance} that the area under the curve of all classes is relatively similar, but COVID-19's AUC is higher than other classes, i.e., $1$.
Figure \ref{Confusion matrices ECOVNet EfficientNet-B5} shows the confusion matrices of the proposed ECOVNet considering the base model of EfiicientNet-B5.
In Figure \ref{Confusion matrices ECOVNet EfficientNet-B5}, it is clear that for COVID-19, the ensemble methods provide much better results than those without ensemble. For balanced and unbalanced test sets, these methods provide results that are $3-4\%$ better than those without ensemble.
However, ECOVNet shows the ability to detect normal and pneumonia chest X-rays, and it provides the same performance while ensemble or no ensemble for the imbalanced test set, although it shows a slightly better performance when classifying the balanced test set with no ensemble.
Finally, we can say that ECOVNet is an eminent architecture for detecting COVID-19 cases from chest X-ray images, because it focuses on distinguishing features that help distinguish COVID-19 from other types (such as normal and pneumonia).

\subsection{Comparison between ECOVNet and the other models}
\label{Comparison between ECOVNet and other models}

Table \ref{Comparison ECOVNet with other methods} shows the comparison between the proposed method and the latest methods from which to detect COVID-19 using chest X-rays, and we have seen that the proposed method is superior to other methods. 
Some previous methods (namely COVID-Net\cite{2020arXiv200309871W}, EfficientNet-B3\cite{luz2020effective}, DeepCOVIDExplainer\cite{Karim2020DeepCOVIDExplainerEC}) used ImageNet weights and the COVIDx data set, however, one of the previous methods, i.e., DeepCOVIDExplainer, also considered two ensemble strategies.
On the other hand, CovXNet\cite{mahmud_rahman_fattah_2020} used an ensemble method and a transfer learning scheme from non-COVID chest X-rays, while retaining training and testing data sets other than COVIDx.
One of the previous methods\cite{luz2020effective} showed comparable performance to our proposed method in terms of accuracy because it can reach $100\%$.
Another method called PDCOVIDNet\cite{chowdhury2020pdcovidnet} achieved an accuracy of $96.5\%$, which lacked by a small margin compared to our proposed method.
As we have observed that the proposed approach consistently exhibits better classification accuracy in different combinations of ensemble with an imbalanced and a balanced set of test data considering a larger number of COVID-19 chest X-rays.
When comparing the results of the two ensemble methods, we observed that the soft ensemble showed impressive results in classifying COVID-19, and the accuracy and recall were both $100\%$.

\begin{table}[!htbp]
\small
\begin{minipage}{\textwidth}
\centering
\caption{Comparison of the proposed ECOVNet with other state-of-the-art methods on COVID-19 detection}
\begin{tabular}{ccccc}
\hline
Method      & Total chest X-rays & \makecell{Precision($\%$)\\(COVID-19)} & \makecell{Recall($\%$)\\(COVID-19)} & Accuracy($\%$) \\ \hline
COVID-Net\cite{2020arXiv200309871W}     &  \makecell{$573$ COVID-19 , $8,066$ Normal,\\$5,559$ Pneumonia} &       $99.0$            & $95.0$      & $94.5$         \\ \hline
\makecell{EfficientNet-B3\cite{luz2020effective}\\(Flat Classification)}         &  \makecell{$183$ COVID-19 , $8,066$ Normal,\\$5,521$ Pneumonia}            &  $\textbf{100}$        &    $96.8$              &   $93.9$       \\ \hline
\makecell{EfficientNet-B3\cite{luz2020effective}\\(Hierarchical Classification)} & \makecell{$183$ COVID-19 , $8,066$ Normal,\\$5,521$ Pneumonia}             &   $\textbf{100}$   &    $80.6$              &      $93.5$    \\ \hline
DeepCOVIDExplainer\cite{Karim2020DeepCOVIDExplainerEC}           & \makecell{$358$ COVID-19 , $8,066$ Normal,\\$5,538$ Pneumonia}             &   $90.4$                  &  $92.7$       & $96.1$         \\ \hline
\multirow{2}{*}{CovXNet\cite{mahmud_rahman_fattah_2020}} & \makecell{$305$ COVID-19 , $305$ Viral Pneumonia,\\$305$ Bacterial Pneumonia}             &       $88.5$              &  $90.3$                &  $89.6$        \\ \cline{2-5} 
                             &  \makecell{$305$ COVID-19 , $305$ Viral Pneumonia,\\$305$ Bacterial Pneumonia + $305$ Normal }            &       $90.8$              &    $89.9$              &  $90.3$        \\ \hline
PDCOVIDNet\cite{chowdhury2020pdcovidnet}          & \makecell{$219$ COVID-19 , $1,341$ Normal,\\$1,345$ Viral Pneumonia}             &   $95.4$                  &  $91.3$       & $96.5$         \\ \hline
\makecell{ECOVNet-Hard Ensemble\footnote{\label{note1}Imbalanced test set}\\(Proposed)}       &  \makecell{$589$ COVID-19 , $8,851$ Normal,\\$6,053$ Pneumonia}           &   $94.1$                  &     $97.0$             &   $96.0$       \\ \hline
\makecell{ECOVNet-Soft Ensemble\textsuperscript{\ref{note1}}\\(Proposed)}       &   \makecell{$589$ COVID-19 , $8,851$ Normal,\\$6,053$ Pneumonia}           &   $92.5$                  &    $\textbf{100}$              &   $96.0$       \\ \hline
\makecell{ECOVNet-Hard Ensemble\footnote{\label{note2}Balanced test set}\\(Proposed)}       &  \makecell{$589$ COVID-19 , $8,851$ Normal,\\$6,053$ Pneumonia}           &   $\textbf{100}$                  &     $97.0$             &   $95.7$       \\ \hline
\makecell{ECOVNet-Soft Ensemble\textsuperscript{\ref{note2}}\\(Proposed)}       &   \makecell{$589$ COVID-19 , $8,851$ Normal,\\$6,053$ Pneumonia}           &   $\textbf{100}$                  &    $\textbf{100}$              &   $\textbf{97.0}$       \\ \hline
\end{tabular}
\label{Comparison ECOVNet with other methods}
\end{minipage}
\end{table}

\subsection{Visualization using Grad-CAM}
\label{Visualization using Grad-CAM}

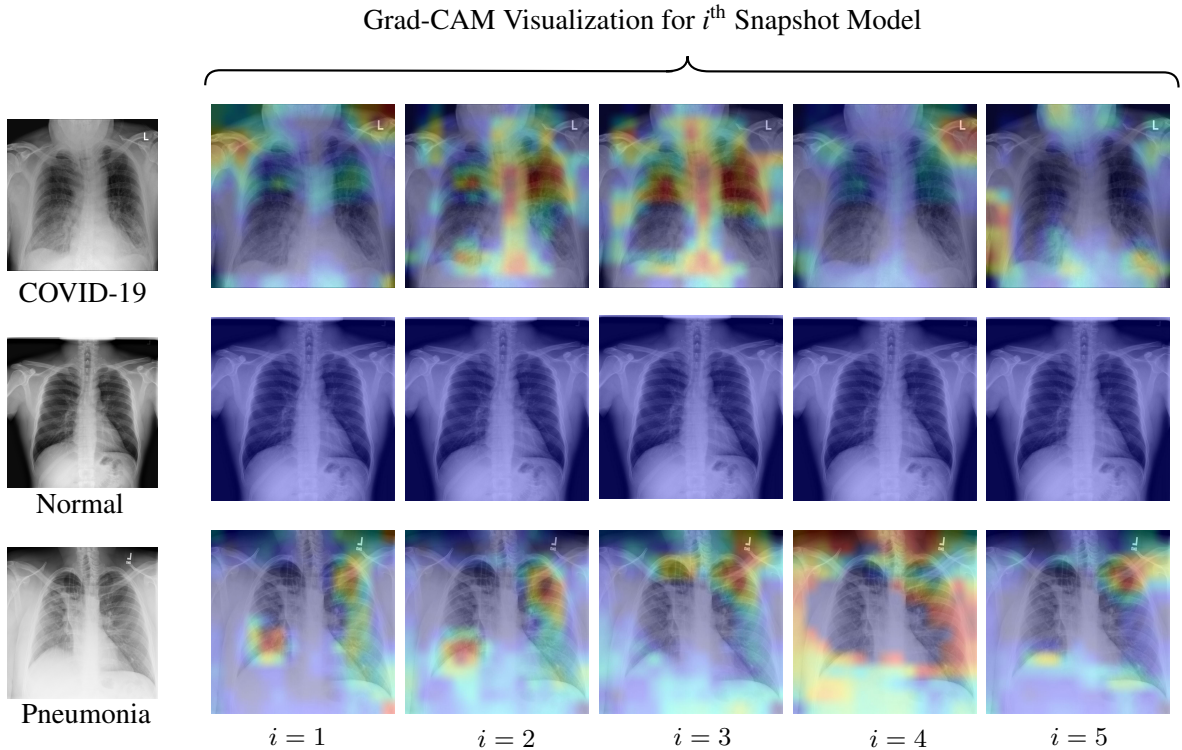
\begin{figure}[!htbp]
\centering
\resizebox{16cm}{!}{

\tikzset{every picture/.style={line width=0.75pt}} 

\begin{tikzpicture}[x=0.75pt,y=0.75pt,yscale=-1,xscale=1]

\draw (96,114) node  {\includegraphics[width=52.5pt,height=52.5pt]{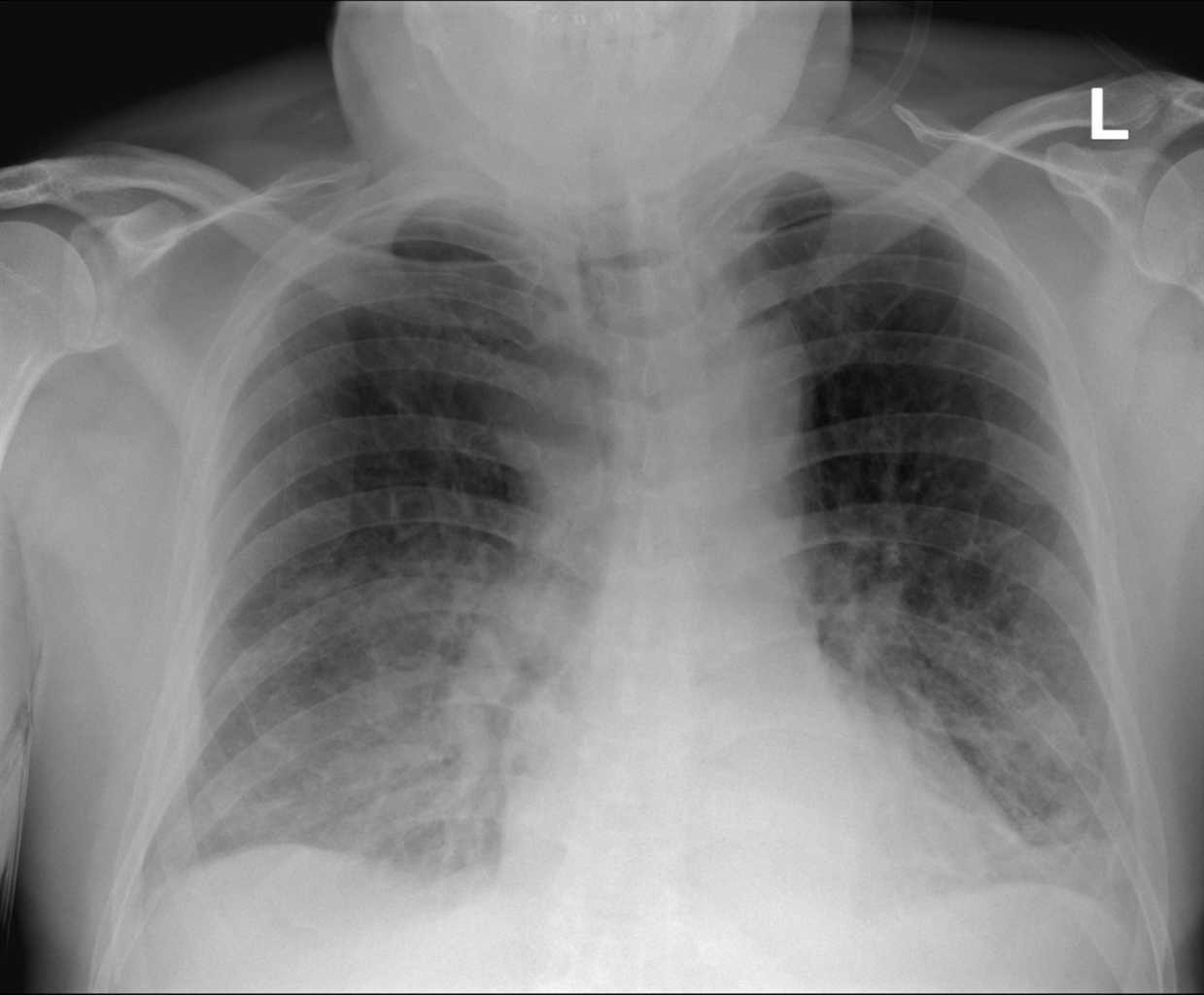}};
\draw (96,215) node  {\includegraphics[width=52.5pt,height=52.5pt]{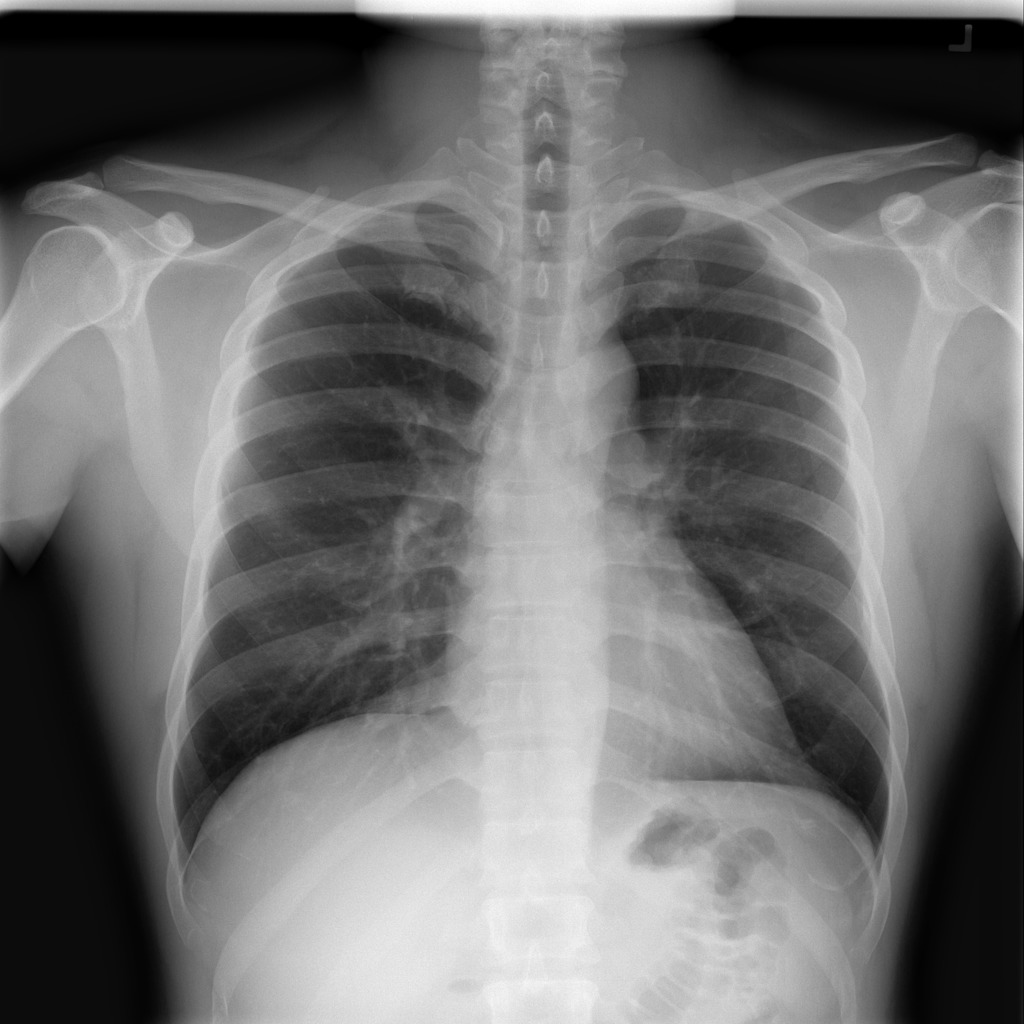}};
\draw (96,313) node  {\includegraphics[width=52.5pt,height=52.5pt]{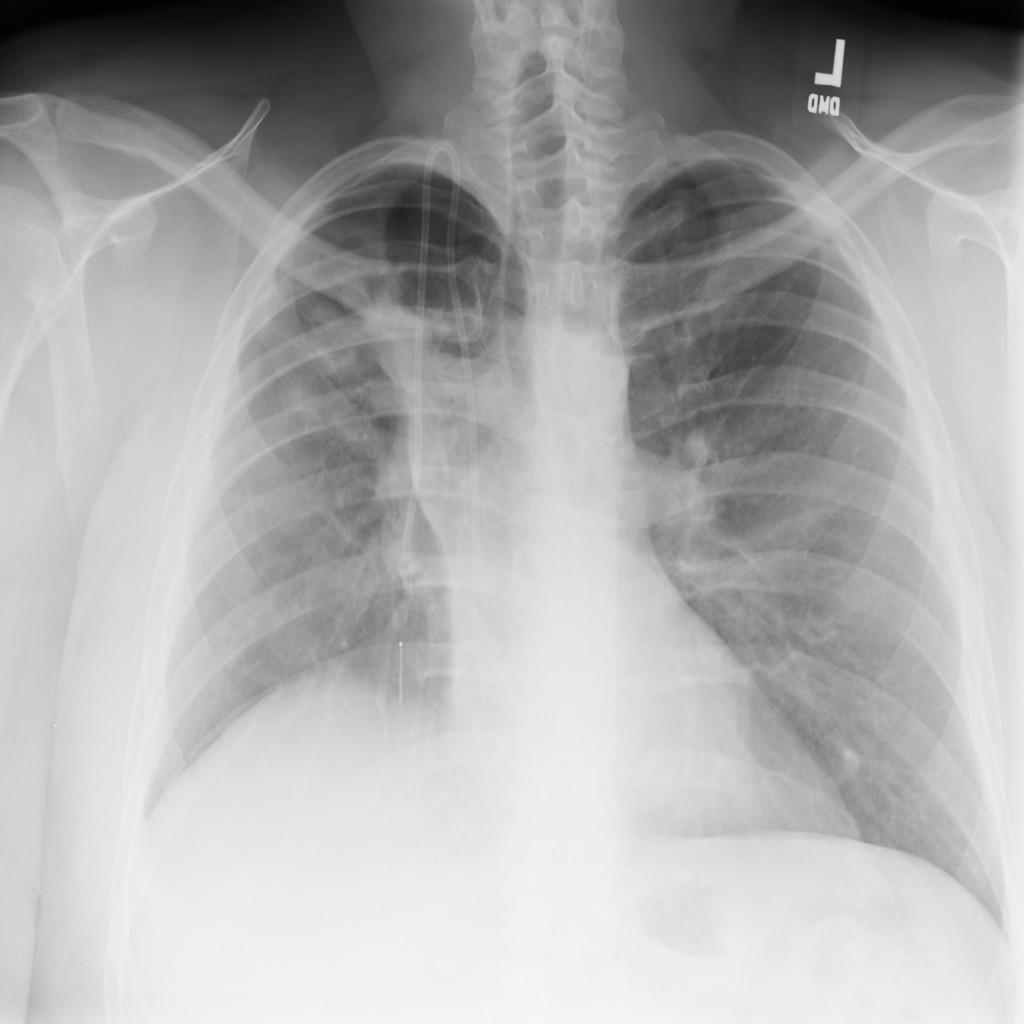}};
\draw (198.5,312.5) node  {\includegraphics[width=63.75pt,height=63.75pt]{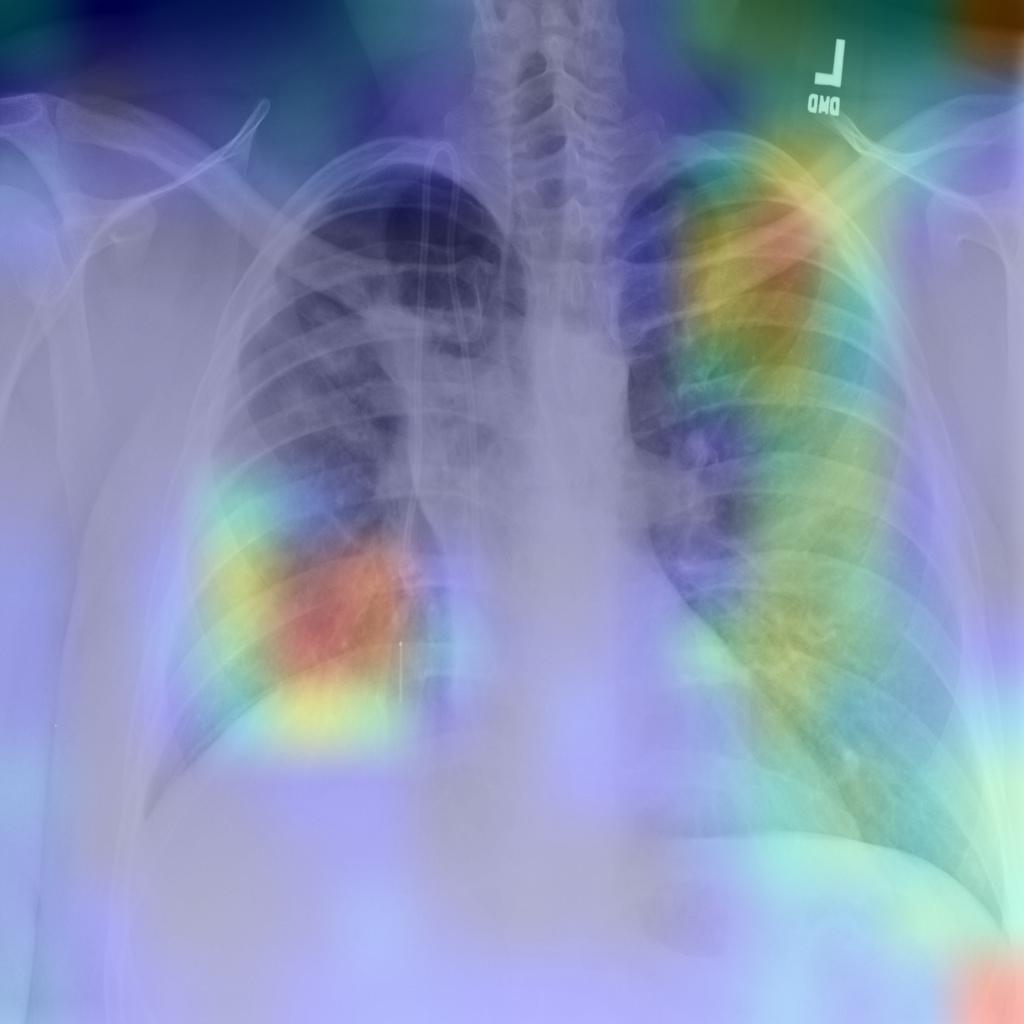}};
\draw (288.5,312.5) node  {\includegraphics[width=63.75pt,height=63.75pt]{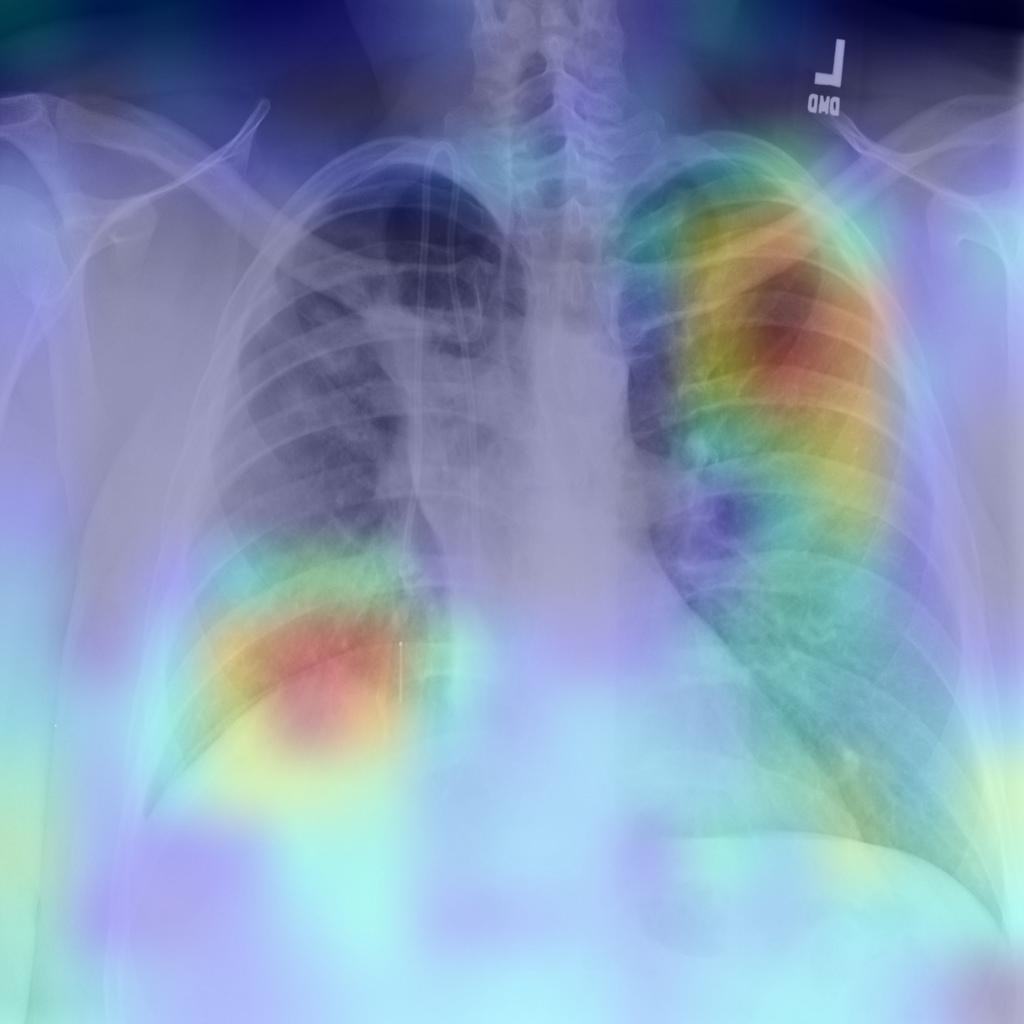}};
\draw (378.5,312.5) node  {\includegraphics[width=63.75pt,height=63.75pt]{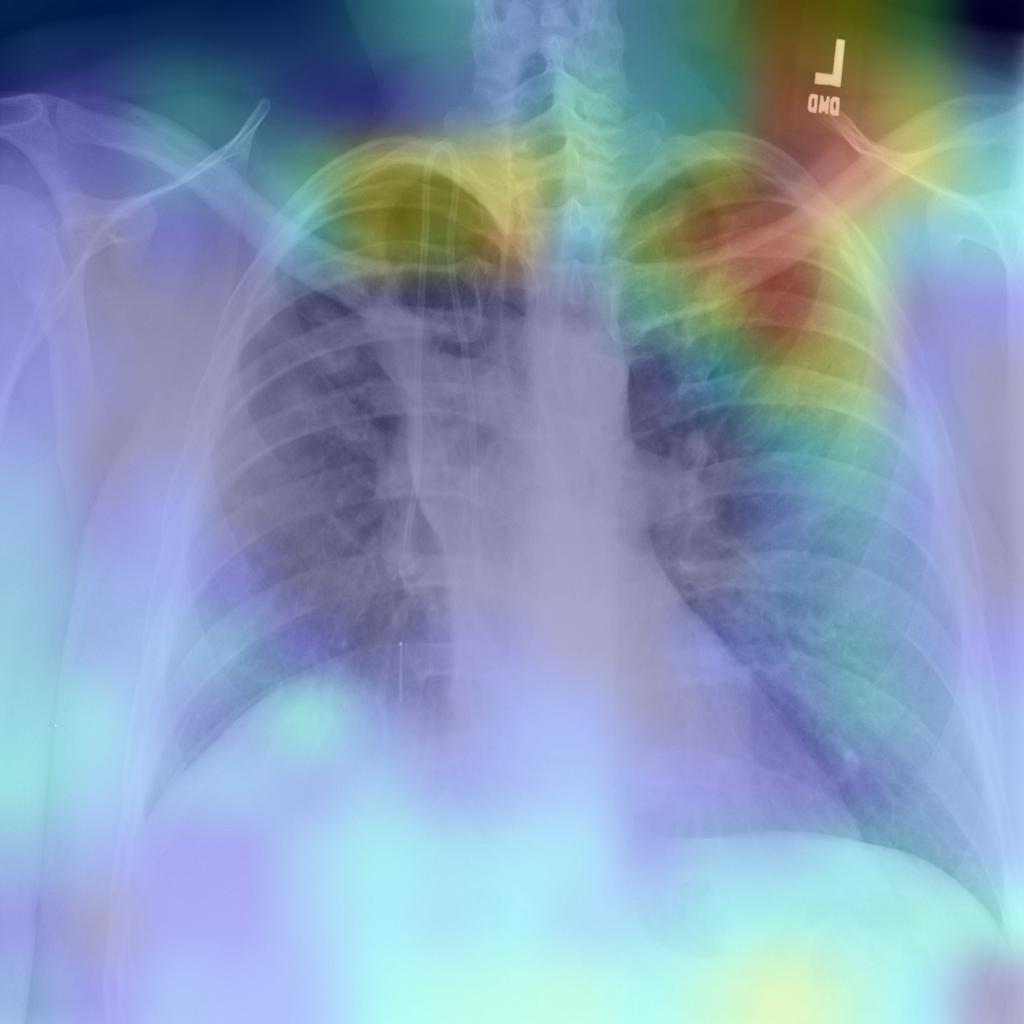}};
\draw (468.5,312.5) node  {\includegraphics[width=63.75pt,height=63.75pt]{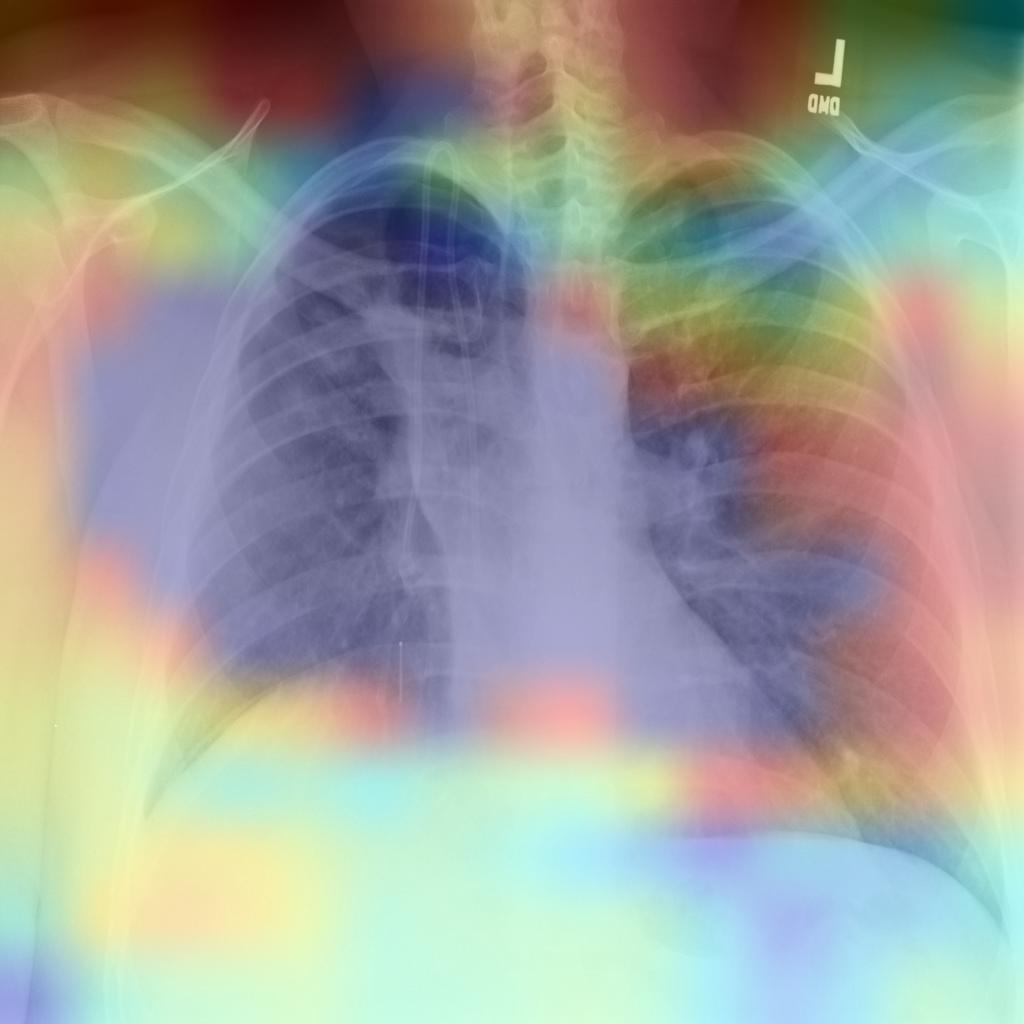}};
\draw (558.5,312.5) node  {\includegraphics[width=63.75pt,height=63.75pt]{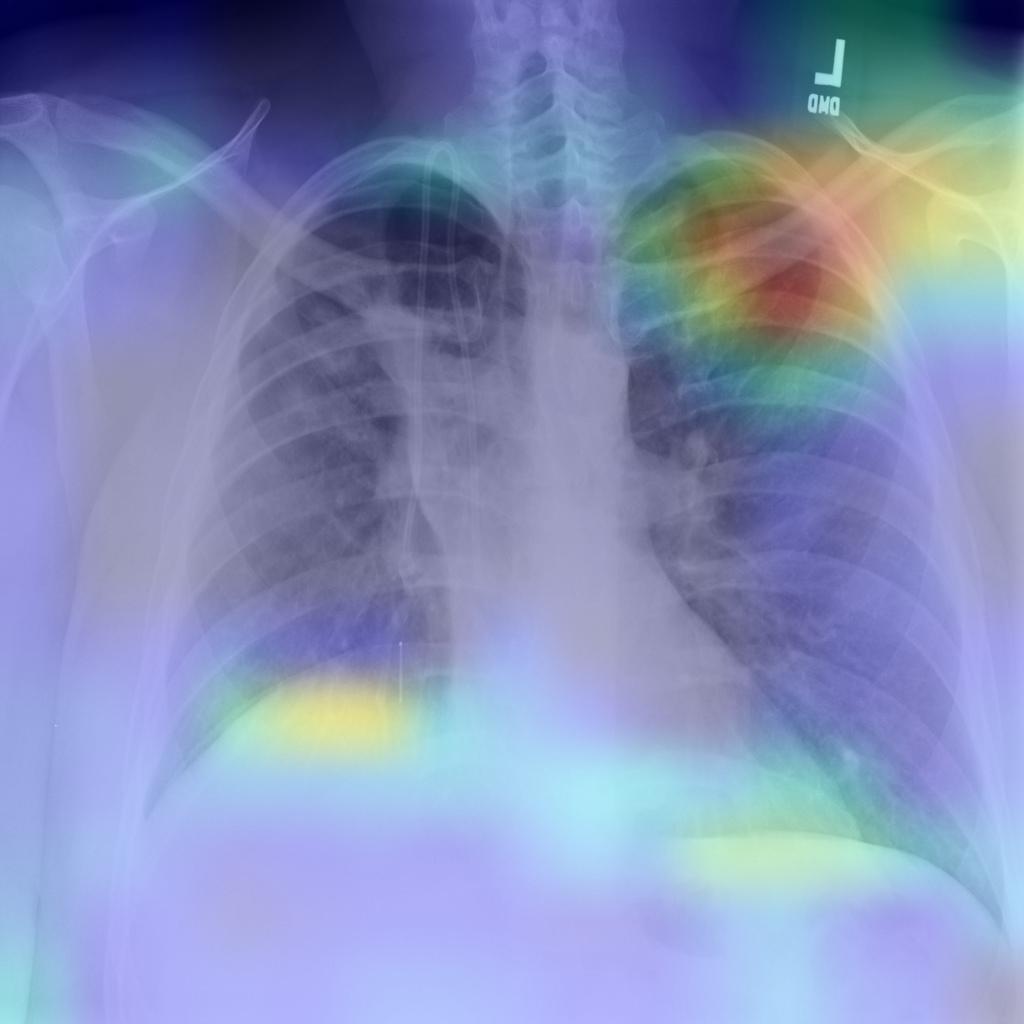}};
\draw (198.5,213.5) node  {\includegraphics[width=63.75pt,height=63.75pt]{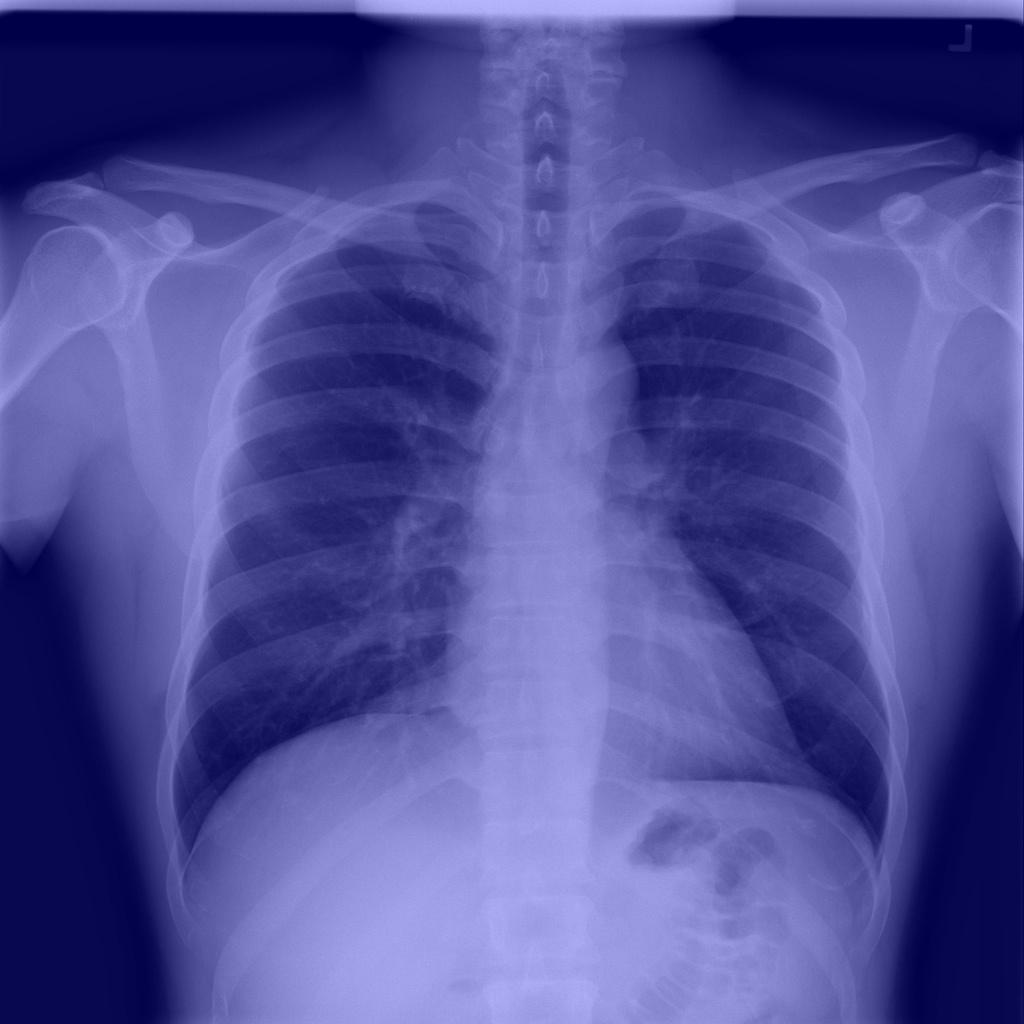}};
\draw (288.5,213.5) node  {\includegraphics[width=63.75pt,height=63.75pt]{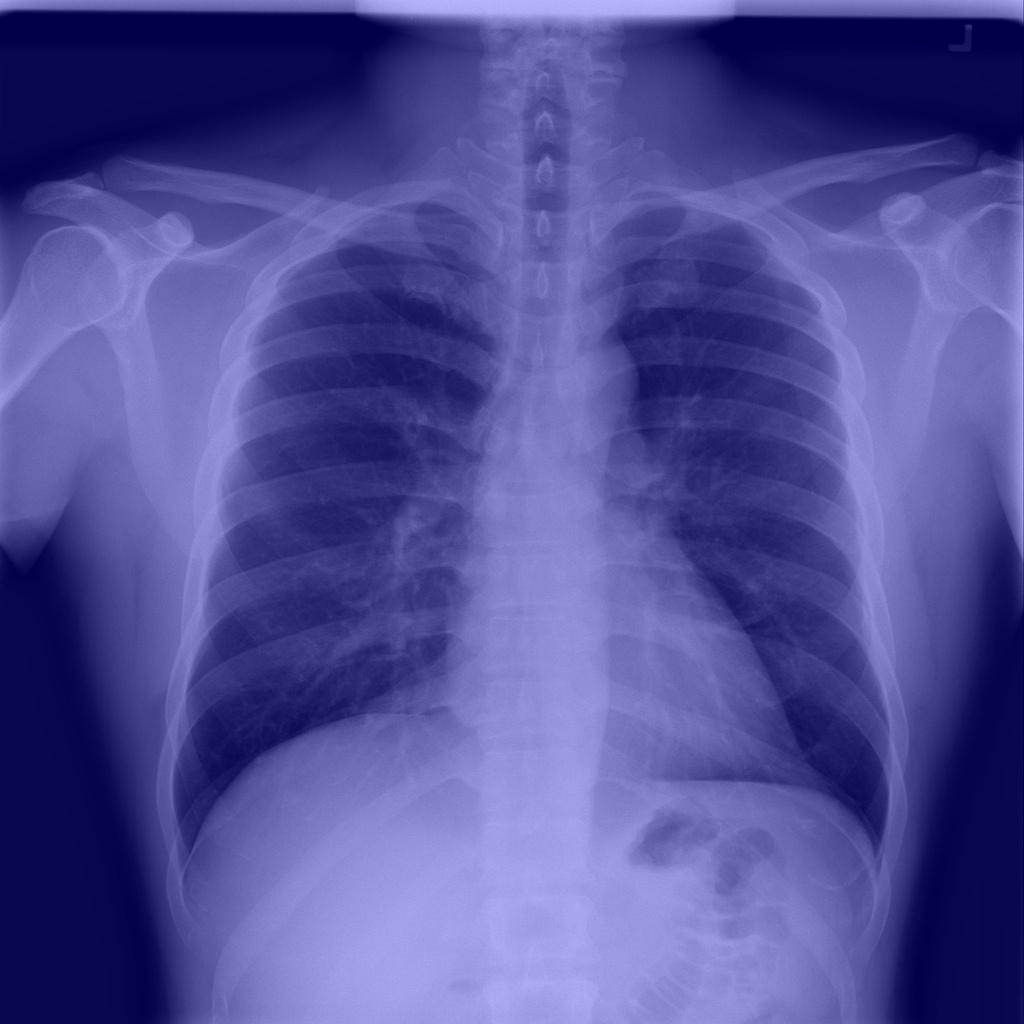}};
\draw (378.5,212.5) node  {\includegraphics[width=63.75pt,height=63.75pt]{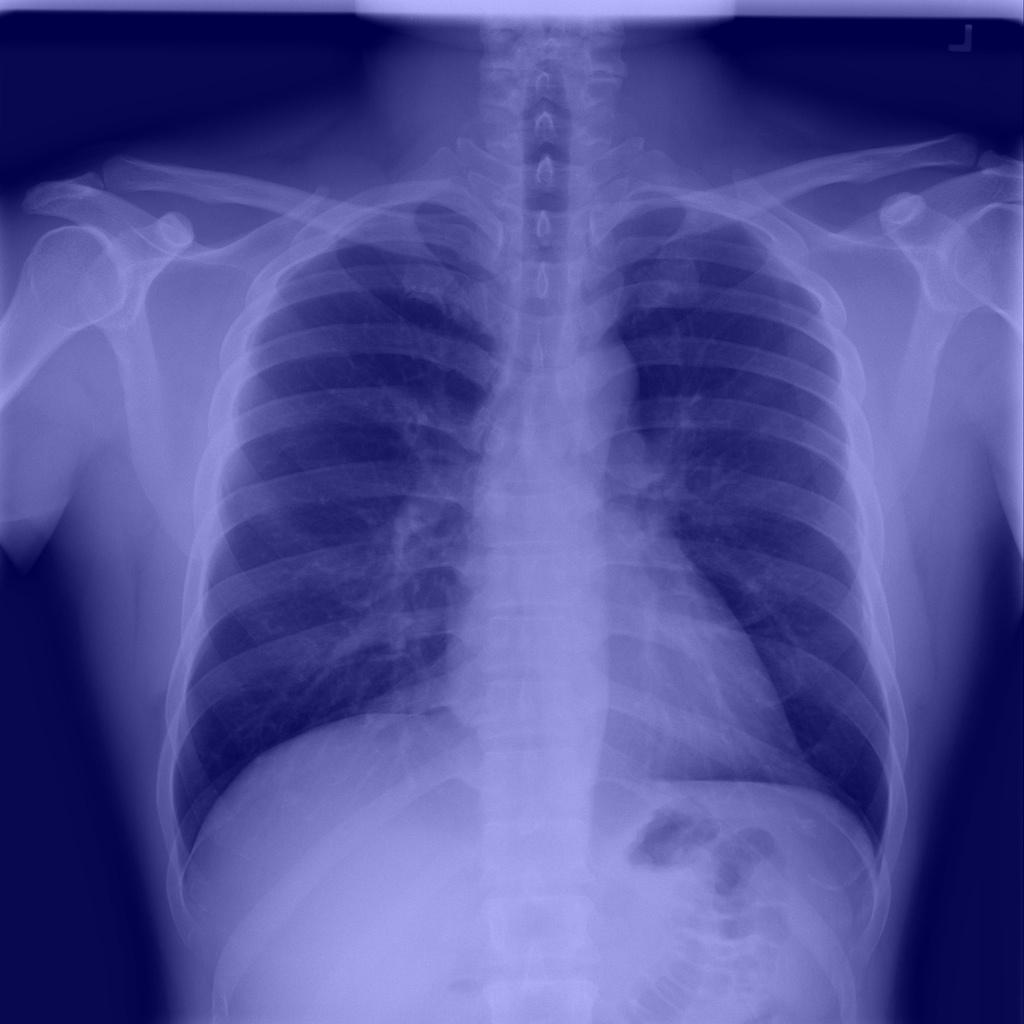}};
\draw (468.5,213.5) node  {\includegraphics[width=63.75pt,height=63.75pt]{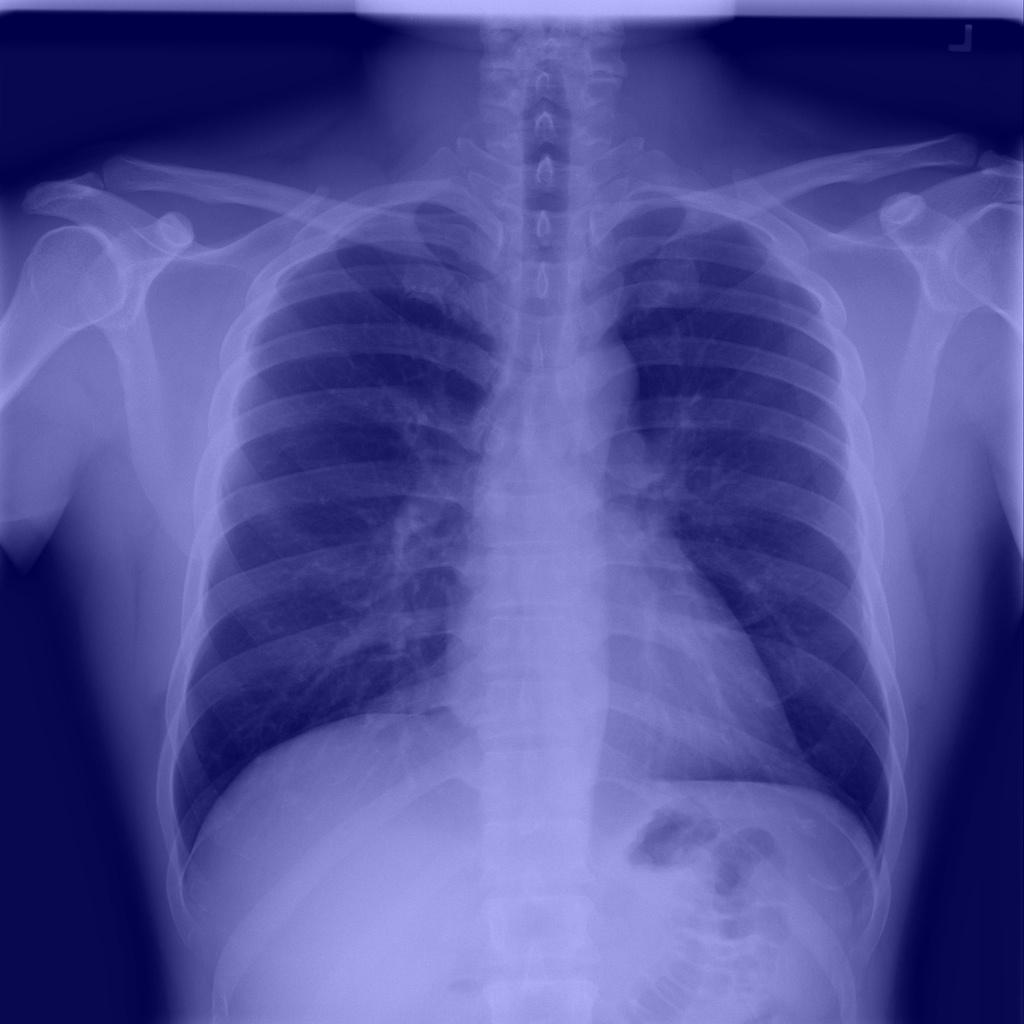}};
\draw (558.5,213.5) node  {\includegraphics[width=63.75pt,height=63.75pt]{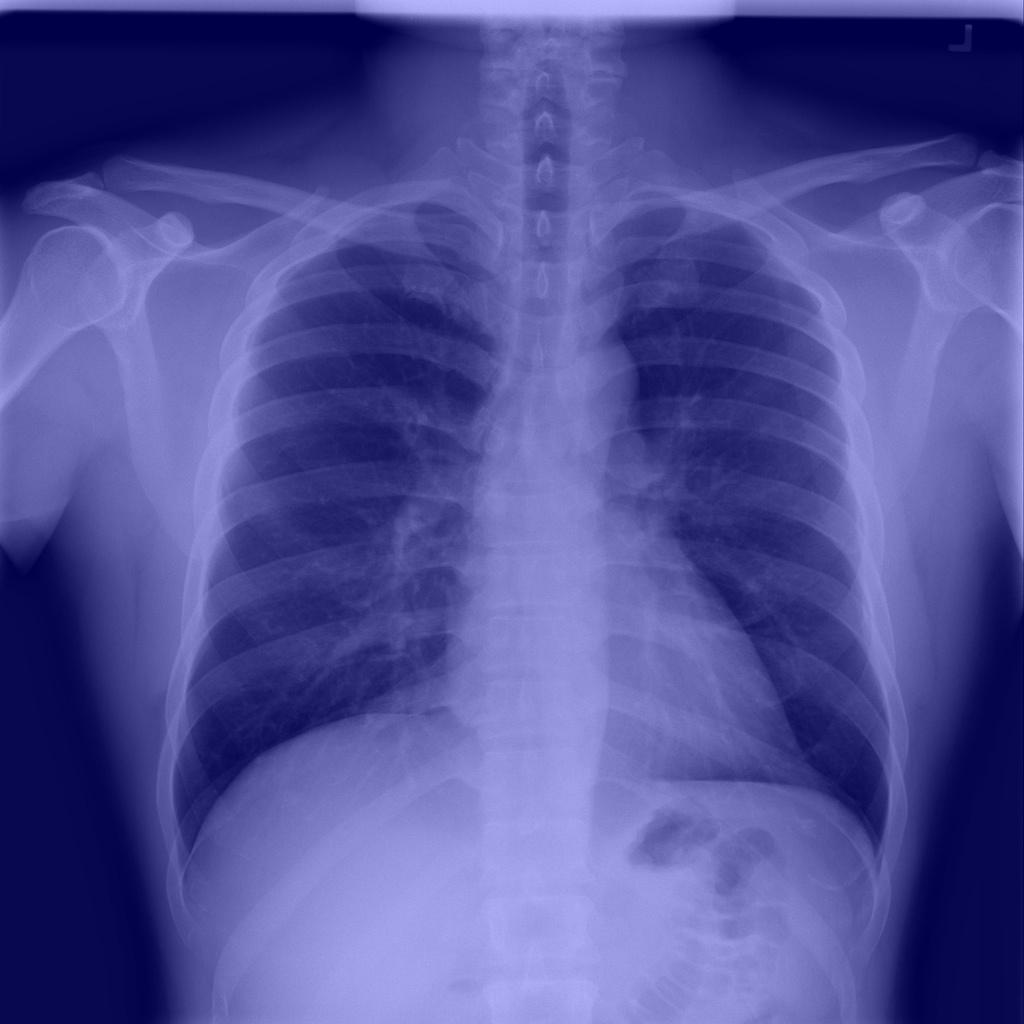}};
\draw (198.5,114.5) node  {\includegraphics[width=63.75pt,height=63.75pt]{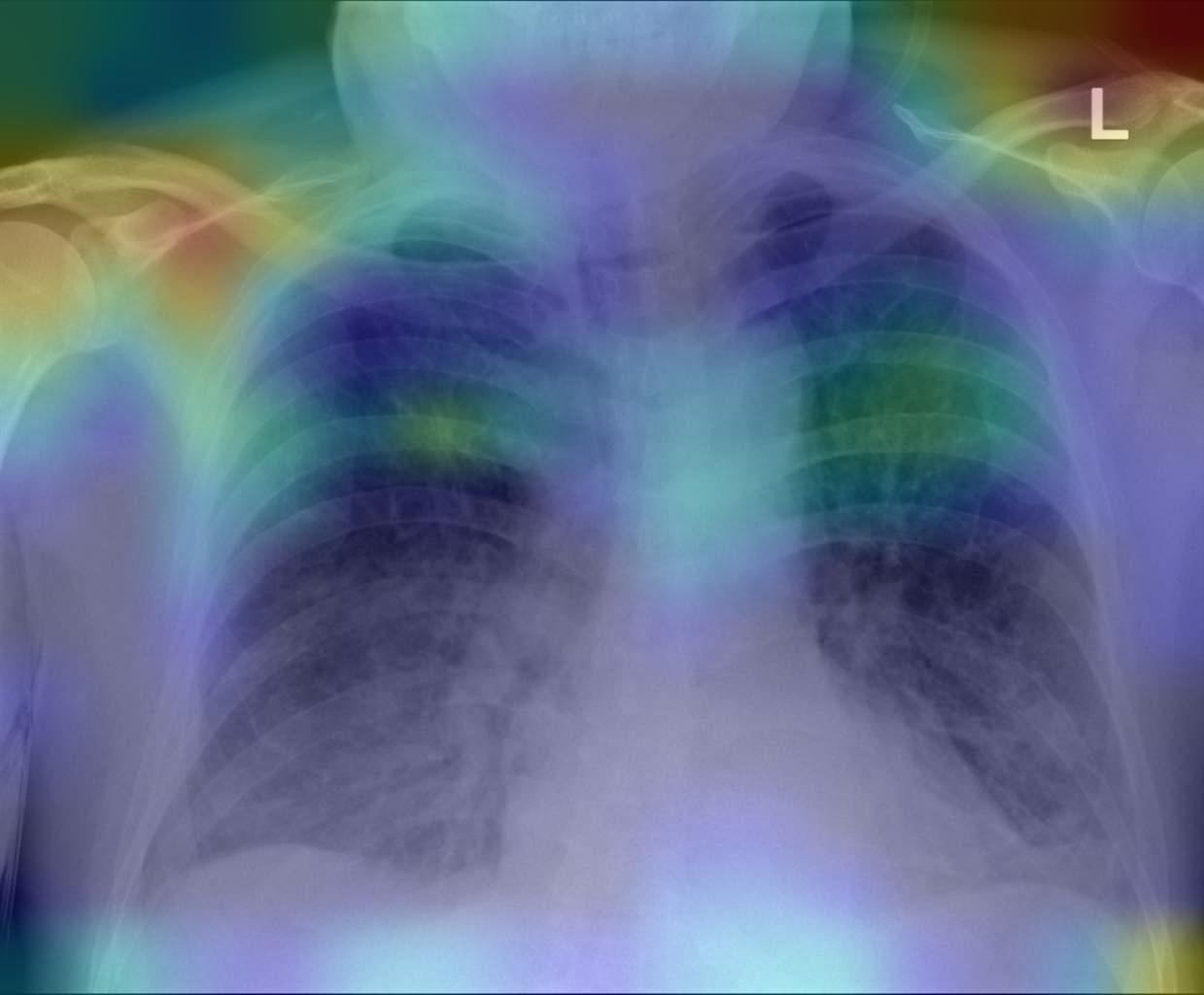}};
\draw (288.5,114.5) node  {\includegraphics[width=63.75pt,height=63.75pt]{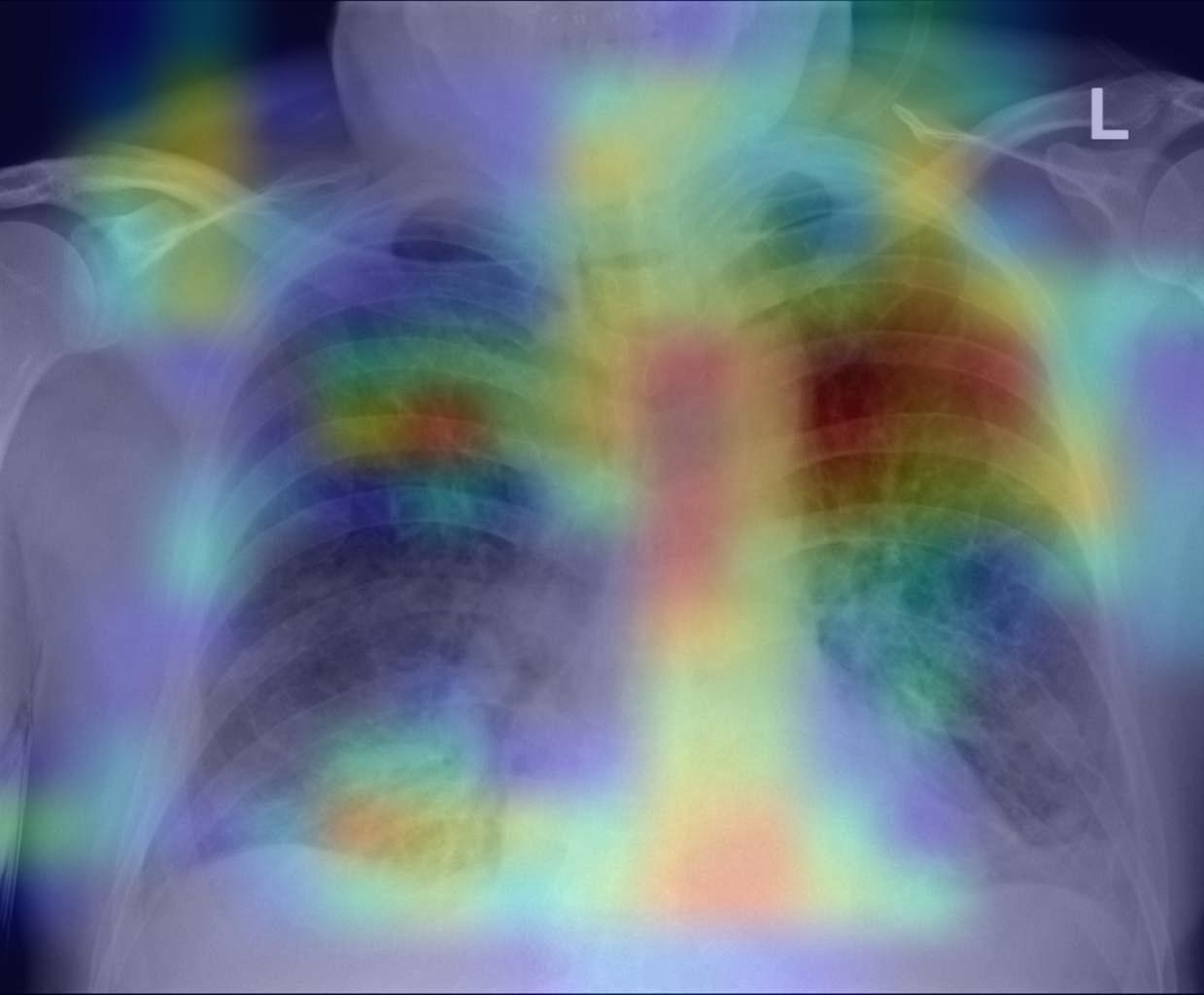}};
\draw (378.5,114.5) node  {\includegraphics[width=63.75pt,height=63.75pt]{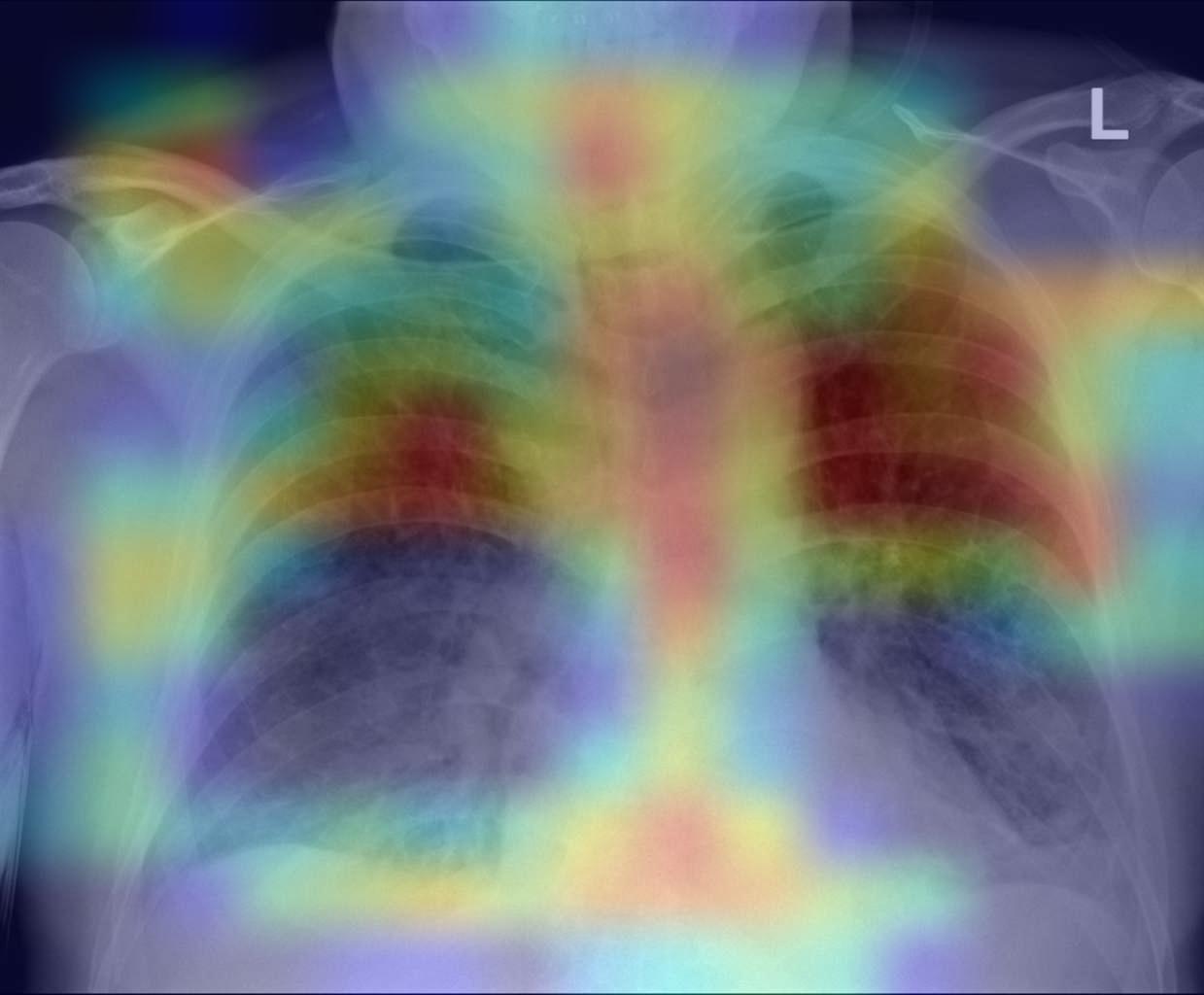}};
\draw (468.5,114.5) node  {\includegraphics[width=63.75pt,height=63.75pt]{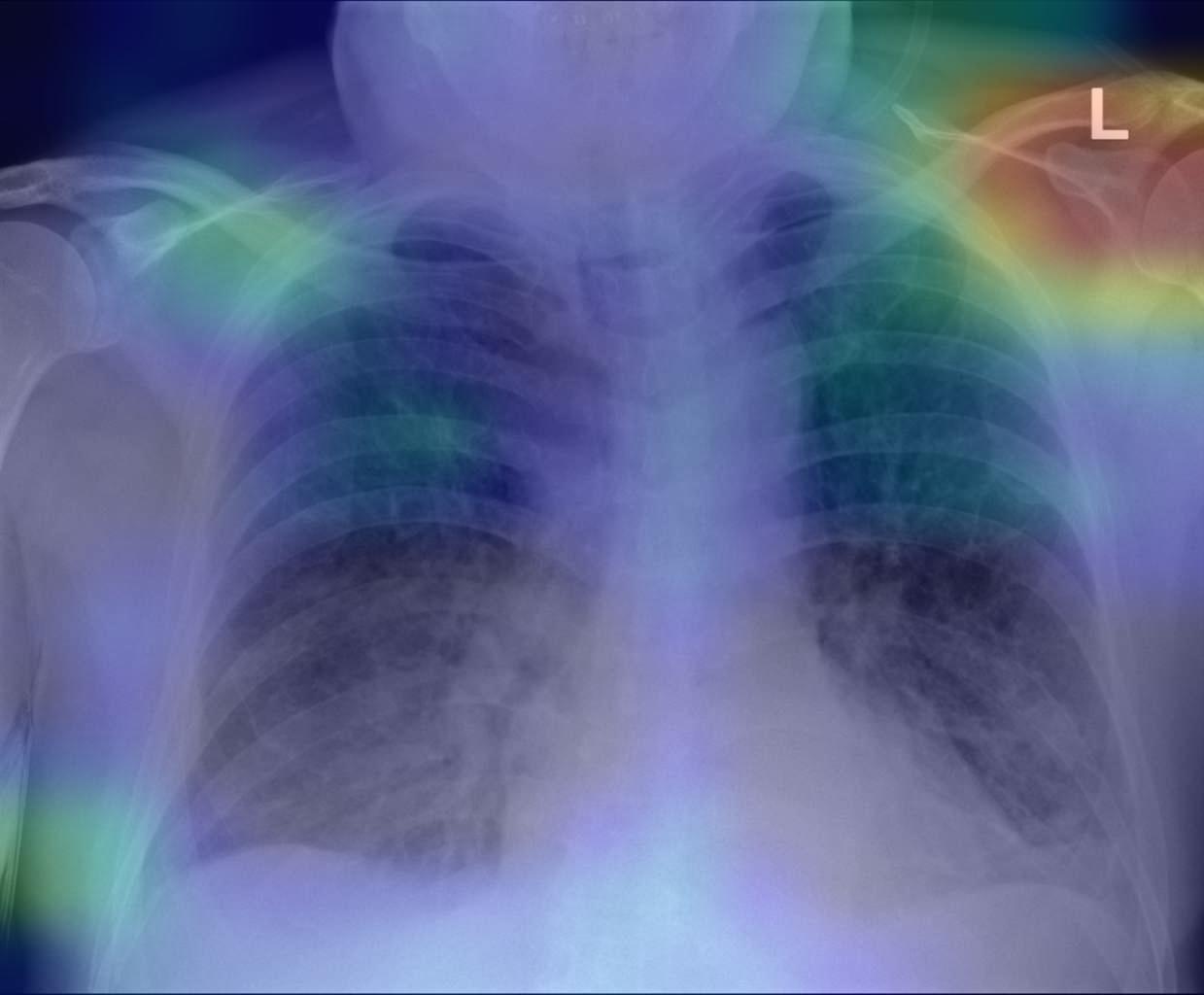}};
\draw (558.5,114.5) node  {\includegraphics[width=63.75pt,height=63.75pt]{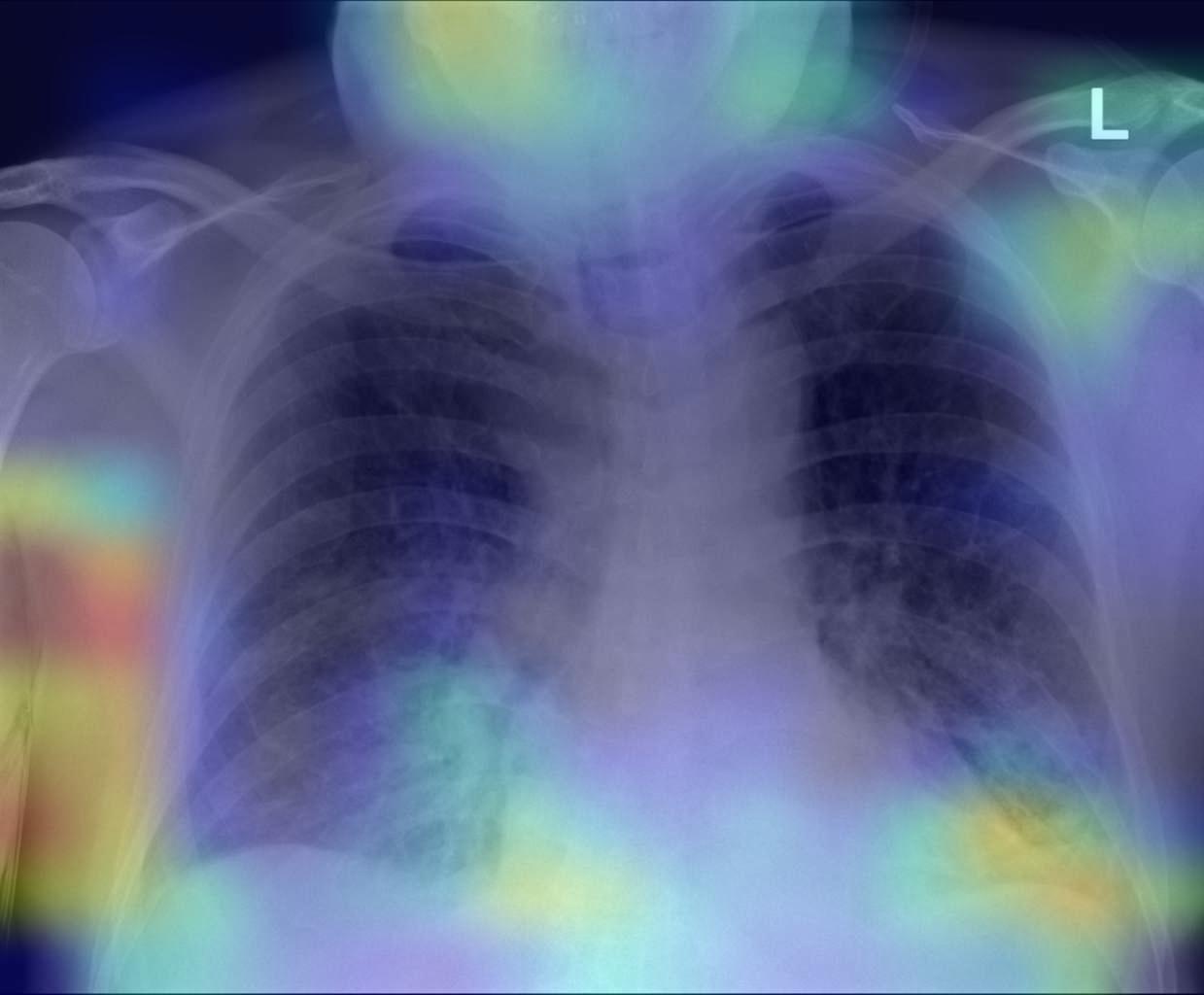}};
\draw   (605,64) .. controls (605.01,59.33) and (602.69,56.99) .. (598.02,56.98) -- (386.99,56.51) .. controls (380.32,56.49) and (376.99,54.15) .. (377,49.48) .. controls (376.99,54.15) and (373.66,56.47) .. (366.99,56.46)(369.99,56.47) -- (160.02,55.99) .. controls (155.35,55.98) and (153.01,58.3) .. (153,62.97) ;

\draw (181,360) node [anchor=north west][inner sep=0.75pt]   [align=left] {\text{{\small {\fontfamily{ptm}\selectfont $i=1$}}}};
\draw (278,361) node [anchor=north west][inner sep=0.75pt]   [align=left] {\text{{\small {\fontfamily{ptm}\selectfont $i=2$}}}};
\draw (369,360) node [anchor=north west][inner sep=0.75pt]   [align=left] {\text{{\small {\fontfamily{ptm}\selectfont $i=3$}}}};
\draw (460,361) node [anchor=north west][inner sep=0.75pt]   [align=left] {\text{{\small {\fontfamily{ptm}\selectfont $i=4$}}}};
\draw (544,360) node [anchor=north west][inner sep=0.75pt]   [align=left] {\text{{\small {\fontfamily{ptm}\selectfont $i=5$}}}};
\draw (65,153) node [anchor=north west][inner sep=0.75pt]   [align=left] {COVID-19};
\draw (73,251) node [anchor=north west][inner sep=0.75pt]   [align=left] {Normal};
\draw (60,349) node [anchor=north west][inner sep=0.75pt]   [align=left] {\begin{minipage}[lt]{54.32560800000001pt}\setlength\topsep{0pt}
\begin{center}
Pneumonia
\end{center}

\end{minipage}};
\draw (225,24) node [anchor=north west][inner sep=0.75pt]   [align=left] {Grad-CAM Visualization for \textit{i}\textsuperscript{th} Snapshot Model};

\end{tikzpicture}

}

\caption{Grad-CAM visualization for the proposed ECOVNet considering the base model EfficientNet-B5. A total of 5 (five) model snapshots were generated during the training process.}

\label{Grad-CAM ECOVNet Snapshot Models}

\end{figure}

In our evaluation, we applied the Grad-CAM visual interpretation method to visually depict the salient areas where ECOVNet emphasizes the classification decision for a given chest X-ray image.
Accurate and definitive salient region detection is crucial for the analysis of classification decisions as well as for assuring the trustworthiness of the results.
In order to locate the salient area, the feature weights with various illuminations related to feature importance are used to create a two-dimensional heat map and superimpose it on a given input image.
Figure \ref{Grad-CAM ECOVNet Snapshot Models} shows the visualization results of locating Grad-CAM using ECOVNet for each model snapshots. 
This salient area locates the area of each category area in the lung that has been identified when a given image is classified as COVID-19 or normal or pneumonia.
As shown in Figure \ref{Grad-CAM ECOVNet Snapshot Models}, for COVID-19, a ground-glass opacity(GGO) occurs along with some consolidation, thereby partially covering the markings of the lungs.
Hence, it leads to lung inflammation in both the upper and lower zones of the lung.
When examining the heat maps generated from the COVID-19 chest X-ray, it can be distinguished that the heat maps created from snapshot 2 and snapshot 3 points to the salient area (such as GGO).
However, in the case of the normal chest X-ray, no lung inflammation is observed, so there is no significant area, thereby easily distinguishable from other classes, i.e., COVID-19 and pneumonia.
As well, it can be observed from the chest X-ray for pneumonia is that there are GGOs in the middle and lower parts of the lungs.
The heat maps generated for the pneumonia chest X-ray are localized in the salient regions with GGO, but for the 4th snapshot model, it appears to fail to identify the salient regions as the heat map highlights outside the lung.
Accordingly, we believe that the proposed ECOVNet provides sufficient information about the inherent causes of the COVID-19 disease through an intuitive heat map, and this type of heat map can help AI-based systems interpret the classification results achieved from the proposed architecture.

\section{Conclusion and Future Work}
\label{Conclusion}

In this paper, we proposed a new modular architecture ECOVNet based on CNN, which can effectively detect COVID-19 with the class activation maps from one of the largest publicly available chest X-ray data set, i.e., COVIDx.
In this work, a highly effective CNN structure (such as the EfficientNet base model with ImageNet pre-trained weights) is used as feature extractors, while fine-tuned pre-trained weights are considered for related COVID-19 detection tasks.
Also, ensemble predictions can improve performance by exploiting the predictions obtained from the proposed ECOVNet model snapshots.
From empirical evaluations, it is observed that the soft ensemble of the proposed ECOVNet model snapshots outperformed the other state-of-the-art methods.
Finally, we performed a visualization study to locate significant areas in the chest X-ray through the class activation map, which is used to classify the chest X-ray into its expected category.
What's more, we believe that our findings will make a useful contribution to the control of COVID-19 infection and the widespread acceptance of automated applications in medical practice.

While this work contributes to reduce the effort of health professional’s radiological assessment, our further plan is to lead this work to design a fully-functional application using guidelines of the design research paradigm~\cite{Miah_Gammack_2014, miah2008ontology}. Such a modern methodological lens could offer further directions both for developing innovative clinical solutions and associative knowledge in the body of relevant literature.
Furthermore, we will spring up a mobile application that can be able to prognosticate whether the disease will become a deadly or not through analyzing a patient’s short term historical chest X-ray pattern if the patient manifests any clinical symptoms related to COVID-19 disease. 
Therefore, this might be a new way to prevent and stop the spread of the COVID-19 pandemic.

\bibliographystyle{unsrt}  
\bibliography{references}  

\end{document}